\documentclass[letterpaper,twocolumn,10pt]{article}
\usepackage{usenix}

\usepackage[available]{usenixbadges}

\usepackage{tikz}
\usepackage{amsmath}

\usepackage{filecontents}

\usepackage{cite}
\usepackage{graphicx}
\usepackage{amsmath}
\usepackage{pifont}
\usepackage{graphicx}
\usepackage{subfigure}
\usepackage{svg}
\usepackage{amssymb}
\usepackage{booktabs}
\usepackage{threeparttable}
\usepackage{tabularx}
\usepackage{threeparttable}
\usepackage{multirow}
\usepackage[table]{xcolor}
\usepackage{colortbl}
\usepackage{xspace}
\usepackage{lipsum} 

\usepackage{algorithm}
\usepackage[noend]{algpseudocode}
\usepackage{tikz}

\usepackage{enumitem}
\setlist[itemize]{leftmargin=1em}

\newcommand{\ProvX}{{\textsc{ProvX}}\xspace}
\newcommand{\SLEUTH}{\textsc{SLEUTH}\xspace}

\newcommand{\Unicorn}{\textsc{Unicorn}\xspace}
\newcommand{\Sigl}{\textsc{Sigl}\xspace}
\newcommand{\ATLAS}{\textsc{ATLAS}\xspace}
\newcommand{\ThreaTrace}{\textsc{ThreaTrace}\xspace}
\newcommand{\ShadeWatcher}{\textsc{ShadeWatcher}\xspace}
\newcommand{\MAGIC}{\textsc{MAGIC}\xspace}
\newcommand{\Kairos}{\textsc{Kairos}\xspace}
\newcommand{\RCAID}{\textsc{R-caid}\xspace}
\newcommand{\FLASH}{\textsc{FLASH}\xspace}
\newcommand{\Orthrus}{\textsc{Orthrus}\xspace}
\newcommand{\Slot}{\textsc{Slot}\xspace}

\newcommand{\upp}[1]{\color[rgb]{0.117, 0.447, 0.999}#1}
\newcommand{\dow}[1]{\color[rgb]{0.753,0,0}#1}
\newcommand{\inc}[1]{\upp\mathbf{\blacktriangle #1\%}}
\newcommand{\dec}[1]{\dow\mathbf{\blacktriangledown #1\%}}

\newcommand*\circledred[1]{\tikz[baseline=(char.base)]{
		\node[shape=circle, draw=red, thick, text=red, inner sep=1.0pt, minimum size=6pt] (char) {#1};
}}

\newcommand*\circledblue[1]{\tikz[baseline=(char.base)]{
		\node[shape=circle, draw=blue, thick, text=blue, inner sep=1.0pt, minimum size=6pt] (char) {#1};
}}

\renewcommand{\footnoterule}{%
	\kern -3pt 
	\hrule width 100pt 
	\kern 2.6pt 
}

\begin{document}
%-------------------------------------------------------------------------------

%don't want date printed
\date{}

\title{Cutting the Fuse: Actionable APT Attack Blocking in
	Provenance-based IDS}

%for single author (just remove % characters)
%\author{
%	Weiheng Wu\textsuperscript{1,3}\thanks{Weiheng Wu and Wei Qiao contributed equally to this work.},
%	Wei Qiao\textsuperscript{1,2},
%	Teng Li\thanks{Corresponding authors: Teng Li.}\textsuperscript{1,2,5},
%	Yebo Feng\textsuperscript{4},
%	Zhuo Ma\textsuperscript{1},
%	Jianfeng Ma\textsuperscript{1},
%	Yang Liu\textsuperscript{4}
%	\\[0.5ex]
%	\textsuperscript{1}School of Cyber Engineering, Xidian University
%	\\
%	\textsuperscript{2}State Key Laboratory of Integrated Services Networks (ISN)
%	\\
%	\textsuperscript{3}Singapore Management University
%	\quad
%	\textsuperscript{4}Nanyang Technological University
%	\\
%	\textsuperscript{5}Songshan Laboratory
%	\\[0.5ex]
%	\small
%	\texttt{wdz478149507@gmail.com,
%		qiaoweiXidian@gmail.com,
%		litengxidian@gmail.com}
%	\\
%	\small
%	\texttt{yebo.feng@ntu.edu.sg,
%		mazhuo@mail.xidian.edu.cn,
%		jfma@mail.xidian.edu.cn,
%		yangliu@ntu.edu.sg}
%}

\author{
	Weiheng Wu\textsuperscript{1}\thanks{Weiheng Wu and Wei Qiao contributed equally to this work.},
	Wei Qiao\textsuperscript{1,3}\hyperlink{HFN.1}{\footnotemark[1]},
	Teng Li\textsuperscript{1,3,4}\thanks{Corresponding authors: Teng Li.},
	Yebo Feng\textsuperscript{2},
	Zhuo Ma\textsuperscript{1},
	Jianfeng Ma\textsuperscript{1},
	Yang Liu\textsuperscript{2}
	\\[0.5ex]
	\textsuperscript{1}School of Cyber Engineering, Xidian University
	\\
	\textsuperscript{2}Nanyang Technological University
%	\quad
	\\
	\textsuperscript{3}State Key Laboratory of Integrated Services Networks (ISN)
	\\
	\textsuperscript{4}Songshan Laboratory
	\\[0.5ex]
	\small
	\texttt{wdz478149507@gmail.com,
		qiaoweiXidian@gmail.com,
		litengxidian@gmail.com}
	\\
	\small
	\texttt{yebo.feng@ntu.edu.sg,
		mazhuo@mail.xidian.edu.cn,
		jfma@mail.xidian.edu.cn,
		yangliu@ntu.edu.sg}
}

\maketitle

\begin{abstract}
Provenance-based Intrusion Detection Systems (PIDS) are a critical line of defense against Advanced Persistent Threats (APTs). However, their practical deployment is crippled by alert flooding; state-of-the-art PIDS focus on detection performance, outputting thousands of low-information, unactionable alerts. These outputs fail to isolate the true cause of the detector's alert and cannot provide the security operations center(SOC) with actionable blocking strategies. 

To address this challenge, we propose \ProvX, a novel APT defense framework designed for blocking attack steps within alerts. Unlike existing works, \ProvX does not aim to merely improve attack detection performance; it is dedicated to transforming unactionable raw alerts into a core edge list for immediate review. Specifically, we introduce counterfactual explanation logic to reduce the alerts and find the minimal structural subset (i.e., the core attack edges) within an alert that determines its malicious prediction. This subset, when perturbed, can subvert the model's original prediction. The framework then applies a staged solidification strategy to enhance the precision and stability of the alert analysis. Ultimately, \ProvX outputs an actionable core edge list for immediate response and blocking of critical attack steps. Experiments on DARPA datasets demonstrate that \ProvX can locate key behaviors within alerts that are highly relevant to real-world attacks, and the identified core edges can flip many detector alerts under the simulated intervention used in our model-level evaluation. Furthermore, we explore and provide a preliminary validation of an alert-feedback enhancement framework, showing that \ProvX's analysis results can guide model optimization in adversarial scenarios.
\end{abstract}

\section{Introduction}
\label{sec:intro}

Advanced Persistent Threats (APTs) pose a severe challenge to modern cybersecurity due to their long-term persistence, multi-stage nature, and high stealth. Traditional defense mechanisms (e.g., firewalls, antivirus) struggle to detect such complex attack chains. To counter this threat, the security community's focus has shifted to the host interior, leveraging operating system audit logs to capture and correlate system activities. In this context, data provenance technology has emerged as one of the most promising solutions. The provenance graph constructs a unified causal graph from discrete log entries (e.g., process creation, file access, network connections), providing security analysts with a structured view to depict the entire scope of an attack. 

\begin{figure}[t]
	\centering
	\includegraphics[width=0.48\textwidth]{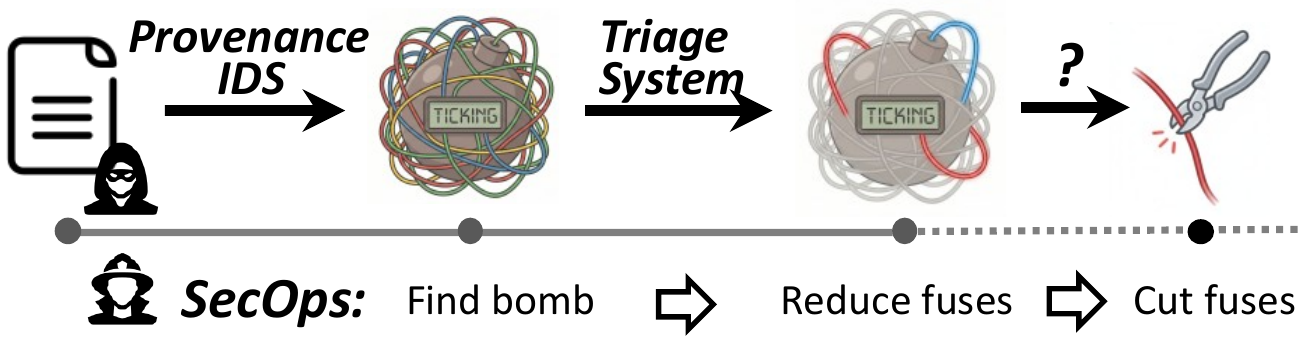}

%	\caption{Statistics on the number of publications on provenance graph detection from 2016 to 2025\cite{ma2016protracer, hossain2017sleuth, ji2017rain, du2017deeplog, ji2018enabling, ma2017mpi,  gao2018saql, pasquier2018runtime, hassan2018towards, kwon2018mci, liu2018towards, milajerdi2019poirot, liu2019log2vec, hassan2019nodoze, milajerdi2019holmes, paccagnella2020logging, hassan2020omegalog, hanunicorn, wang2020you, yang2020uiscope, hassan2020tactical, alsaheel2021atlas, han2021sigl, han2021deepaid, zeng2021watson, yu2021alchemist, fang2022back, zeng2022palantir, king2022euler, xu2022depcomm, zengy2022shadewatcher, yang2023prographer, ding2023case, mukherjee2023evading, dong2023we, altinisik2023provg, goyal2023sometimes, inam2023sok, jia2024magic, li2023nodlink, sekar2024eaudit, cheng2024kairos, rehman2024flash, goyal2024r, qiao2024slot, jiang2025orthrus, bilotsometimes, wang2024incorporating}.}
\caption{The status quo of APT defense in SecOps. }
	% \Description{A figure illustrating four scenes.}
	\label{fig:bomb}
\end{figure}

The rise of provenance technology has spurred the development of provenance-based intrusion detection systems (PIDS). Early PIDS relied primarily on rule-based methods\cite{hassan2020tactical, hossain2017sleuth, milajerdi2019holmes} or statistics-based methods\cite{liu2018towards, hassan2020we}. While these methods could achieve low false-positive rates, they were highly dependent on prior knowledge, struggled to discover unknown attack patterns, and often performed poorly in real-world environments with voluminous and noisy log data.
In recent years, the research focus has clearly shifted to learning-based methods\cite{hanunicorn, pei2016hercule, shen2018tiresias,alsaheel2021atlas, wang2022threatrace, li2021hierarchical, zengy2022shadewatcher, cheng2024kairos, rehman2024flash, goyal2024r, qiao2024slot}, especially Graph Neural Networks (GNNs). The powerful graph representation learning capabilities of GNNs make them naturally suited for processing complex provenance graph structures.

While the research community has achieved excellent performance in attack detection tasks, even exceeding 99\% accuracy, through the design of various graph embedding and aggregation strategies, industry security operations (SecOps) are simultaneously plagued by a problem: alert fatigue. To achieve high recall and capture all potential threats, these systems are designed to be highly sensitive, inevitably leading to an overwhelming flood of alerts, the vast majority of which are unactionable noise. As shown in Figure \ref{fig:bomb}, while a PIDS can successfully identify that a "bomb" is ticking within the host, it leaves analysts with a tangled mess of thousands of potential "fuses"—discrete system events—without any guidance on which ones are critical. Our experience indicates that a typical enterprise SOC is often inundated with thousands of daily alerts, far exceeding the operational capacity of a ten-person team, which can effectively process only 300.

This order-of-magnitude gap between detection and response forces analysts to ignore the vast majority of alerts, allowing real threats to be buried in the noise. While existing triage works (e.g., NoDoze\cite{hassan2019nodoze}) attempt to reduce the number of \textit{fuses} through statistical refinement, they often fail to pinpoint the exact choke point required to disarm the threat. Therefore, we argue that the next critical challenge for the PIDS domain is no longer just improving detection rates, but rather providing analysts with immediately actionable attack blocking solutions—identifying the minimal, precise "wire" that must be cut to neutralize the attack chain.

To address these challenges, we propose \ProvX, an APT alert-refinement framework designed to synthesize blocking-oriented response guidance. We use mitigation in a model-relative sense: \ProvX identifies edges whose simulated removal flips the detector's prediction. Unlike existing PIDS that focus solely on detection, \ProvX aims to provide analysts with candidate choke points within alerted provenance subgraphs. Drawing inspiration from the potential of explainability, we introduce interventional reasoning and formalize the defense task as a minimal intervention problem. This involves pinpointing the critical \textit{fuses} (i.e., the minimal set of system edges) whose simulated removal changes the detector's judgment, thereby revealing the structural evidence most necessary to the alert. Furthermore, we explore a closed-loop enhancement framework, leveraging these blocking-oriented results to guide the optimization of the underlying detector.

In terms of design, \ProvX's core idea is to transform the discrete problem of searching for attack choke points into a continuous optimization task. Specifically, \ProvX learns a numerically differentiable blocking mask to simulate the process of an analyst cutting specific attack fuses within the provenance graph. This process is guided by a dual-objective optimization: first, a prediction flip loss, which drives the model to find the structural intervention required to change the detector's malicious prediction; second, a mask distance loss, which ensures the operational intervention cost is minimized by localizing only the most critical edges. Finally, \ProvX employs a staged solidification strategy to lock in confident judgments on important interactions, enhancing the stability and clarity of the identified choke points. Ultimately, this framework provides a high-confidence, actionable blocking list on top of foundational GNN detectors.

We evaluate \ProvX's mitigation efficacy on the authoritative DARPA TC E3 and DARPA OpTC datasets. These real-world attack scenarios provide realistic attack scenarios for evaluating core-edge localization and model-level blocking efficacy. Our experiments show that \ProvX achieves a recall rate of up to over 70\% in pinpointing ground-truth attack steps. More importantly, we introduce the Mitigation Efficacy Rate (MER) to evaluate whether the recommended core-edge intervention flips the detector's alert under a controlled structural-attribute intervention, achieving an average MER of 45.74\%. We quantitatively define the cost-benefit trade-off for security operations and demonstrate that \ProvX significantly outperforms state-of-the-art PIDS and statistical triage tools. Finally, we validate \ProvX as a forensic analyzer for adversarial evasion, exploring a feedback loop that uses identified evasion structures to harden the detector against future attacks.
Our contributions are as follows:
\begin{itemize}
\item[$\bullet$]We propose \ProvX, a novel APT defense framework. Its goal is not merely attack detection, but to refine massive alerts into an actionable core edge list for attack mitigation.
\item[$\bullet$]\ProvX applies interventional reasoning based on counterfactual logic to identify the most critical system interactions within a detected threat, and formalize attack mitigation as a minimal intervention optimization task.
\item[$\bullet$]We design and validate a closed-loop feedback framework that integrates detection, mitigation, and forensic feedback, that can enhance the underlying detector's robustness against adaptive attacks.
\item[$\bullet$]We evaluate \ProvX on authoritative APT datasets across key dimensions, including localization accuracy, mitigation efficacy, and SecOps cost-benefit performance. Comprehensive comparisons against SOTA PIDS and statistical triage algorithms prove \ProvX's superiority in generating high-value, low-cost blocking strategies.
\end{itemize}
\section{Background and Motivation}
\label{sec:background_motivation}
\subsection{Related work}
\subsubsection{System Provenance}
System provenance technology is widely adopted in academia to conduct attack investigations on large-scale host audit logs(Table \ref{otherwork}). 
%Specifically, by constructing audit logs into a Provenance Graph, we obtain a more intuitive and information-rich representation of the system's execution process. This allows security analysts to perform causal analysis on system activities, thereby enabling intrusion detection and security event attribution.
Research on Provenance-based Intrusion Detection Systems (PIDS) has roughly progressed through three phases. Early PIDS relied primarily on rule-based methods\cite{hassan2020tactical, hossain2017sleuth, milajerdi2019holmes} or statistics-based methods\cite{liu2018towards, hassan2020we}. While these approaches achieved low false-positive rates in specific scenarios by strictly adhering to expert knowledge, predefined attack templates, or event frequencies, their performance was highly dependent on expert guidance. This dependency caused them to sacrifice the detection capability for unknown or potential threats, making them ill-suited for the complexity and variability of APT attacks.

Consequently, the research focus has shifted to learning-based methods\cite{hanunicorn, pei2016hercule, shen2018tiresias,alsaheel2021atlas, wang2022threatrace, li2021hierarchical, zengy2022shadewatcher, cheng2024kairos, rehman2024flash, goyal2024r, qiao2024slot}, particularly Graph Neural Networks (GNNs). The powerful graph representation learning capabilities of GNNs are naturally suited for processing complex provenance graph structures, and their effectiveness in detecting APTs has been demonstrated. A significant body of work has focused on using GNNs and various custom GNN-based models to learn embeddings for downstream detection tasks. Although these methods can achieve good detection performance by understanding the structural and semantic information of audit logs, high false-positive rates and unactionable alerts remain critical bottlenecks. These issues prevent security analysts from effectively localizing the true attack and forming actionable investigation results. Therefore, recent work\cite{cheng2024kairos, rehman2024flash, qiao2024slot} has trended towards performing post-detection attack reconstruction to enhance comprehensibility for security analysts, even if this only reveals highly aggregated attack groups.

\begin{table}[t]
	
	\centering
	\resizebox{0.48\textwidth}{!}{
		\renewcommand{\arraystretch}{1.3}
		\begin{threeparttable}
			\begin{tabular}{ccccc}
				\hline
				\textbf{System} & \textbf{Time} & \textbf{Encoder} & \textbf{Decoder} & \textbf{Granularity}            \\ \hline
				\SLEUTH\cite{hossain2017sleuth}        & 2017          & -                & -                & Path   \\ \hline
				%				\NODOZE\cite{hassan2019nodoze}\tnote{2}         & 2019          & -                & -                & Graph   \\ \hline
				\Unicorn\cite{hanunicorn}& 2020 & - & - & Graph \\ \hline
				\Sigl\cite{han2021sigl}           & 2021          & Graph LSTM       & MLP              & Graph   \\ \hline
				\ATLAS\cite{alsaheel2021atlas}          & 2021          & LSTM+CNN         & -       & -   \\ \hline
				\ThreaTrace\cite{wang2022threatrace}     & 2022          & GraphSAGE        & -                & Node \\ \hline
				\ShadeWatcher\cite{zengy2022shadewatcher}   & 2022          & TransR+GNN       & Inner Product    & Edge \\ \hline
				\MAGIC\cite{jia2024magic}          & 2024          & GAT              & GAT+MLP          & Node \\ \hline
				\Kairos\cite{cheng2024kairos}         & 2024          & TGN              & MLP              & Graph   \\ \hline
				\RCAID\cite{goyal2024r}          & 2024          & GAT              & -                & Path   \\ \hline
				\FLASH\cite{rehman2024flash}          & 2024          & GraphSAGE        & XGBoost          & Node   \\ \hline
				\Orthrus\cite{jiang2025orthrus} & 2025 & Graph-Transformer & MLP & Node \\ \hline
				\Slot\cite{qiao2024slot}           & 2025          & GCN+RL           & MLP              & Node   \\ \hline
			\end{tabular}
%			\begin{tablenotes}
%				\item[1] SLEUTH builds rules based on typical or specific APT patterns and matches these rules to detect.
%				%				\item[2] \NODOZE measures anomalies based on the frequency of related events.
%				\item[2] \Unicorn summarizes provenance graphs into fixed-size sketches using a Weisfeiler-Lehman Kernel and HistoSketch, then uses K-medoids clustering to model normal system states.
%			\end{tablenotes}
		\end{threeparttable}
	}
	\caption{Analysis of representative provenance-based detection studies in recent years.}
	\label{otherwork}
\end{table}

\subsubsection{Alert Fatigue and Triage}
Despite high detection performance, modern PIDS face a critical operational hurdle: Alert Fatigue. A typical enterprise SOC may receive over 17,000 alerts per week, with over 51\% of them considered false positives (and some studies \cite{hassan2019nodoze} citing this figure as high as 60-80\%)\cite{tariq2025alert}. This volume creates an order-of-magnitude gap between detection and the limited response capacity of security analysts.

To mitigate this, triage tools like NoDoze\cite{hassan2019nodoze} and DepImpact\cite{fang2022back} were developed to rank alerts based on statistical heuristics, such as event frequency or path weights. However, these methods are often ill-equipped to handle Living-off-the-Land (LotL) attacks, where attackers use legitimate system commands that mimic benign statistical signatures. Consequently, these triage efforts merely reduce the number of leads without identifying the truly critical nodes in the attack chain.

\subsubsection{Actionability in SecOps: The Missing Link}
The ultimate goal of security operations is not just to detect a "ticking bomb" but to effectively disarm it. However, a significant actionability gap persists between state-of-the-art PIDS and practical defense. While current systems may accurately flag a subgraph as malicious, they often output thousands of unprioritized entities, leaving analysts without a clear blocking strategy.

As highlighted in the SoK by Inam et al.\cite{inam2023sok} , current anomaly-based detection offers poor actionable explanations. The study suggests that an effective solution must isolate anomalous activity to a minimal set of paths to be useful for investigators. We argue that the next frontier for PIDS lies in interventional actionability—transitioning from post-hoc explanation to providing a prescriptive defense scheme that pinpoints the candidate choke points that should be prioritized for blocking-oriented response.

\begin{figure}[h] % 建议为figure环境加上位置参数，如htbp
	\centering 
	\subfigure[Embedding distribution analysis of pure semantic encoding.]{ % 建议为子图添加标题
		\includegraphics[width=0.20\textwidth]{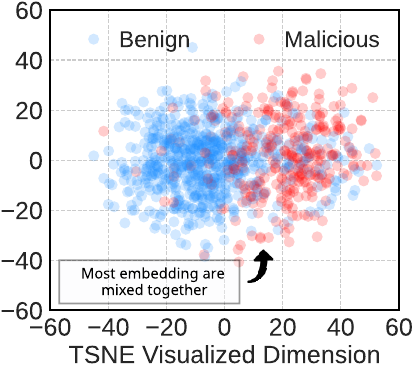}
		\label{fig:onlySemantics}
	}%
	% \hspace*{-5.5pt}%
	\subfigure[Embedding distribution analysis of structural coding]{
		\includegraphics[width=0.20\textwidth]{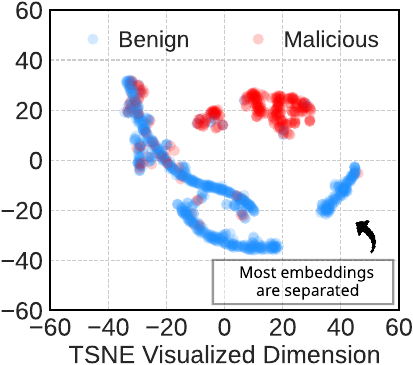}
		\label{fig:onlyGcn}
	}%
	\caption{Embedding distributions of benign and malicious subgraph patterns.}
	\label{fig:mov_key_ob}
\end{figure}

\subsection{Motivation of \ProvX}
\subsubsection{Key Observation}

We first examine which signal is most reliable for distinguishing malicious behavior from benign behavior. By decomposing a complete APT attack provenance graph, we can obtain several smaller subgraphs that represent specific attack patterns or benign behaviors. As illustrated in Figure \ref{fig:mov_key_ob}, by comparing benign and malicious subgraphs in terms of both semantic and structural similarity, we arrive at a key observation: subgraphs related to different attacks exhibit clear structural similarities to one another, and the structural properties of these attack subgraphs also show significant differences from those of benign subgraphs. In contrast, semantic attributes do not offer this clear distinguishing capability.

%This key observation clearly demonstrates that topological structure is more robust and reliable than the semantics of system entities, and is a more effective distinguishing signal. 
In provenance graphs, causal dependency structure is a necessary condition for the existence of attack chains. Attackers can forge filenames (semantics), but it is difficult to forge causal paths (structure) without breaking the attack chain. Mature cybersecurity frameworks, such as Kill Chain and MITRE ATT\&CK\cite{ATTCK}, also confirm this finding, describing the interrelated and predictable structural dependencies between attack strategies and techniques. Therefore, our framework focuses on precisely locating these key structural bottlenecks in attribution graphs to generate a list of actionable APT mitigation and prevention measures.

\begin{figure}[t]
	\centering
	\includegraphics[width=0.48\textwidth]{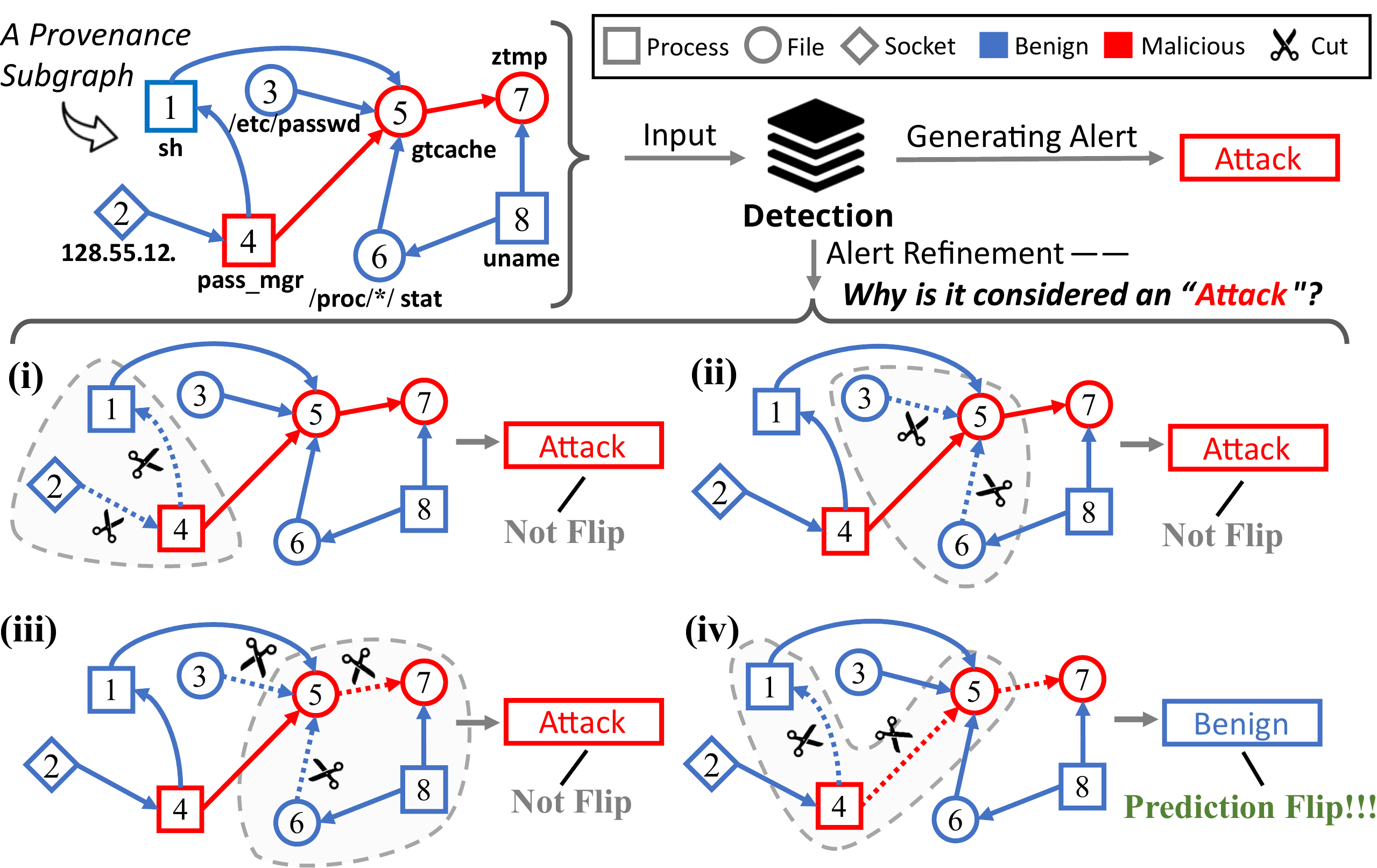}
	%\vspace{-0.1in}
%	\caption{An attack example from DARPA E3 Trace, (i)-(iv) represent different defense attempt in different counterfactual reasoning, which will lead to two results, the model predicts \textit{\textbf{Not Flip}} or \textit{\textbf{ Prediction Flip}}. \textit{\textbf{ Prediction Flip}} results will help to block attack.}
	\caption{A simplified attack example from DARPA TC E3 Trace. Panels (i)--(iv) show different simulated interventions under counterfactual reasoning, leading to either \textit{\textbf{Not Flip}} or \textit{\textbf{Prediction Flip}} outcomes in the detector. A prediction flip indicates that the selected edges are important to the detector's malicious judgment and can guide analyst review of candidate blocking points.}
	% \Description{A figure illustrating four scenes.}
	\label{fig:mov}
%	\vspace{-0.1in}
\end{figure}

\subsubsection{Attack Blocking for Provenance Detection}

We use the APT attack shown in the Figure \ref{fig:mov} as a driving example for illustration. In this scenario, the attacker infiltrates the target host via a browser extension named \textcolor{gray}{\texttt{pass\_mgr}} and subsequently executes the \textcolor{gray}{\texttt{gtcache}} program. \textcolor{gray}{\texttt{gtcache}} then establishes communication with the attacker's Command and Control (C\&C) server. Its primary objective is to scan the host for sensitive information and operate on the \textcolor{gray}{\texttt{/ztmp}} directory, preparing for subsequent attack stages. For clarity in the subsequent discussion, each system entity in the figure has been numbered.

Taking the attack case above as an example, some existing studies are limited to graph analysis, outputting only graph-level alerts\cite{cheng2024kairos, hanunicorn}. This coarse granularity poses a significant challenge for security analysts: it only informs the analyst that ``this graph is problematic" without revealing which specific internal structures triggered the alert. Simply put, the analyst is unable to discern the root cause of the alert. Furthermore, while subsequent research has begun to focus on node-level\cite{wang2022threatrace, rehman2024flash, qiao2024slot} or edge-level\cite{zengy2022shadewatcher} detection tasks, the effort to capture all attacks has resulted in a high volume of false positives, directly leading to alert flooding. Our focus is therefore on how to automatically refine a suspicious alerted graph down to its core components—the parts that analysts can immediately review, remediate and block the attack.

We observe that, in provenance-based alerts, attack blocking and counterfactual reasoning share a useful structural analogy. Security operations seeks low-cost interventions that disrupt suspicious dependencies (e.g., terminating a process, deleting a file) , while counterfactual reasoning seeks minimal structural perturbations that flip the detector's prediction. \ProvX uses this analogy to derive model-level blocking guidance, without equating prediction flipping with guaranteed real-world attack disruption.
Accordingly, we employ counterfactual logic to refine alerts. Our objective is to investigate: what minimal structural modifications (simulating interventions) to the original alerted graph would cause the detection system to reverse its prediction to benign?
Therefore, we adopt a counterfactual explanation approach to explore various hypothetical structural changes (as shown in (i), (ii), (iii), and (iv) in Fig. \ref{fig:mov}). For instance:

\begin{itemize}
	\item(i) After deleting edges \circledblue{2}-\circledred{4} and \circledred{4}-\circledblue{1} (representing partial interaction between \textcolor{darkgray}{\texttt{pass\_mgr}} and \textcolor{darkgray}{\texttt{sh}}), the alert is still triggered. This suggests that the operation of \textcolor{darkgray}{\texttt{pass\_mgr}} on sh represented by \circledred{4}-\circledblue{1} may have malicious tendencies, but its individual effect is not the root cause for the system triggering the alert.
	
	\item(ii) and (iii) explore other simulated interventions, blocking different sets of interactions such as \circledblue{3}-\circledred{5} or \circledblue{6}-\circledred{5}. These attempts also yield comparable results, failing to flip the system's prediction. This demonstrates that, like the intervention in (i), these targeted changes are insufficient on their own to resolve the alert.
	
%	\item However, in (iv), if both edge \circledred{4}-\circledblue{1} (interaction between \textcolor{darkgray}{\texttt{pass\_mgr}} and \textcolor{darkgray}{\texttt{sh}}) and edge \circledred{4}-\circledred{5} (execution of \textcolor{darkgray}{\texttt{gtcache}} by \textcolor{darkgray}{\texttt{pass\_mgr}}) are deleted, the detection system flips the predicted outcome to benign, successfully block this attack.
	\item However, in (iv), if both edge \circledred{4}-\circledblue{1} (interaction between \textcolor{darkgray}{\texttt{pass\_mgr}} and \textcolor{darkgray}{\texttt{sh}}) and edge \circledred{4}-\circledred{5} (execution of \textcolor{darkgray}{\texttt{gtcache}} by \textcolor{darkgray}{\texttt{pass\_mgr}}) are deleted, the detection system flips the predicted outcome to benign, indicating that these edges are critical to the detector's malicious judgment.
\end{itemize}

%This counterfactual result reveals the critical significance of the \textcolor{darkgray}{\texttt{pass\_mgr}} executing gtcache operation in the security determination, and also indicates that the interaction between \textcolor{darkgray}{\texttt{pass\_mgr}} and \textcolor{darkgray}{\texttt{sh}} is likewise an important part of the malicious topological structure learned by the detection system.
This result reveals that the minimal flipping edge set \circledred{4}-\circledblue{1} and \circledred{4}-\circledred{5} is precisely the critical basis on which the detector triggers the alert. \ProvX automatically refines a large, complex alert into a concise, high-relevance core edge list (Figure \ref{fig:mov}(iv)).
%At the operational level, it pinpoints two critical choke points: preventing \textcolor{darkgray}{\texttt{pass\_mgr}} from accessing sh and from launching gtcache would result in the \textcolor{darkgray}{\texttt{gtcache}} process never being created. Consequently, all of \textcolor{darkgray}{\texttt{gtcache}}'s subsequent malicious behaviors—including \textit{communicating with the C\&C server} and \textit{scanning for sensitive information}—would be preemptively blocked.
At the operational level, these edges become natural candidates for analyst validation: preventing \textcolor{darkgray}{\texttt{pass\_mgr}} from accessing sh and from launching gtcache may stop the \textcolor{darkgray}{\texttt{gtcache}} process from being created, thereby disrupting subsequent behaviors such as \textit{communicating with the C\&C server} and \textit{scanning for sensitive information}. This list helps analysts reason about the alert's likely choke points, and achieve a more thorough and comprehensible analysis of the alert's root cause and led to block this attack.
%This list can help security analysts to achieve a more thorough and comprehensible analysis of the alert's root cause and led to block this attack.
% \input{designgoal}
\section{Problem Formalization}
\label{formalization}

\subsection{Definition}
We define key designs and notations for the task. In a SOC deployment, an alert may originate from a provenance-based detector or from other security signals, such as endpoint alarms, network indicators, or analyst investigation. These alerts can be mapped to relevant host audit events and expanded into suspicious provenance subgraphs for downstream analysis. In this paper, we instantiate this setting with graph-level PIDS alerts, but the formalization only requires an alert-associated provenance subgraph as input. Assume we have $N$ such provenance subgraphs $\{G_1, G_2, G_3, \dots, G_N\}$, where each subgraph $G_k$ corresponds to a ground-truth label $Y_k \in \{0, 1\}$. Here, 0 indicates that all system entities within the subgraph are benign, while 1 indicates that the subgraph contains at least one stage of an APT attack.

We use a GNN-based detector $f(\cdot)$ to classify these provenance subgraphs. During testing, the trained detector predicts each subgraph $G_k$ and outputs its predicted label $\hat{Y}_k$. \ProvX focuses on triggered alerts, i.e., subgraphs predicted as malicious by the detector. The goal of \ProvX is to support attack blocking by converting such an alert into a minimal core edge set that analysts can prioritize for response.

Since a provenance graph is an abstraction of host-level behavior, \ProvX first solves a model-level blocking problem: it asks which simulated edge-level interventions on the alert graph are sufficient to flip the detector's malicious prediction. For an alerted subgraph $G_k$, we generate a counterfactual subgraph $\tilde{G}_k$ by perturbing a subset of edges. If the detector's prediction changes, i.e., $f(\tilde{G}_k) \neq f(G_k)$, the perturbed edges are considered critical to the detector's original alert. Our objective is to find the smallest such structural perturbation, so that the resulting core edge set is concise enough for analyst review and possible host-level enforcement. The mathematical definition is as follows:
\begin{equation} \label{eq:my_optimization}
	\begin{aligned}
		& \min_{\tilde{G}_k}d(\tilde{G}_k, G_k), \\
		& \text{s.t.,} \ \  \hat{\tilde{Y}}_k\neq \hat{Y}_k.
	\end{aligned}
\end{equation}
where $d(\tilde{G}_k, G_k)$ measures the structural difference between the original alert and its counterfactual version, and $\hat{\tilde{Y}}_k$ is the detector's prediction on $\tilde{G}_k$.

This objective defines blocking at the detector level. A feasible solution means that the selected edges are necessary for the current PIDS model to maintain its malicious judgment. It does not by itself prove that the corresponding host-level enforcement action is always safe, feasible, or sufficient to disrupt the real attack. In deployment, the resulting core edge list should therefore be treated as analyst-prioritized blocking guidance rather than an autonomous blocking policy.

\subsection{Threat Model and Analysis Principle}
In our threat model setting, we assume that attackers will leave traceable behavioral patterns and attack traces in the audit logs of the target system. These traces enable the security model we construct to effectively distinguish between malicious attack activities and normal benign system behaviors. Similar to other related research \cite{hanunicorn, milajerdi2019holmes, milajerdi2019poirot, wang2020you}, we presume that the data collected by audit logging tools is complete and acquired on top of a Trusted Computing Base (TCB). This assumption ensures the reliability and authenticity of the provenance graphs we construct \cite{pasquier2017practical, bates2015trustworthy}.  We acknowledge that this assumption may not hold in all deployments: attackers may tamper with logging pipelines or exploit blind spots in audit collection. Under such incomplete logging, \ProvX can only reason over the observed provenance graph.

Considering the increasing prevalence of adversarial attacks in the current field of network security, we also carefully evaluate the performance changes of the core edge localization algorithm proposed in this paper when facing the attack mode that has been obfuscated or adversarially modified. The relevant discussion and experimental results will be presented in detail in the experimental Section \ref{eval} of this paper.

\section{The \ProvX Framework}
\label{detection}

\begin{figure}
	\centering
	\includegraphics[width=0.49\textwidth]{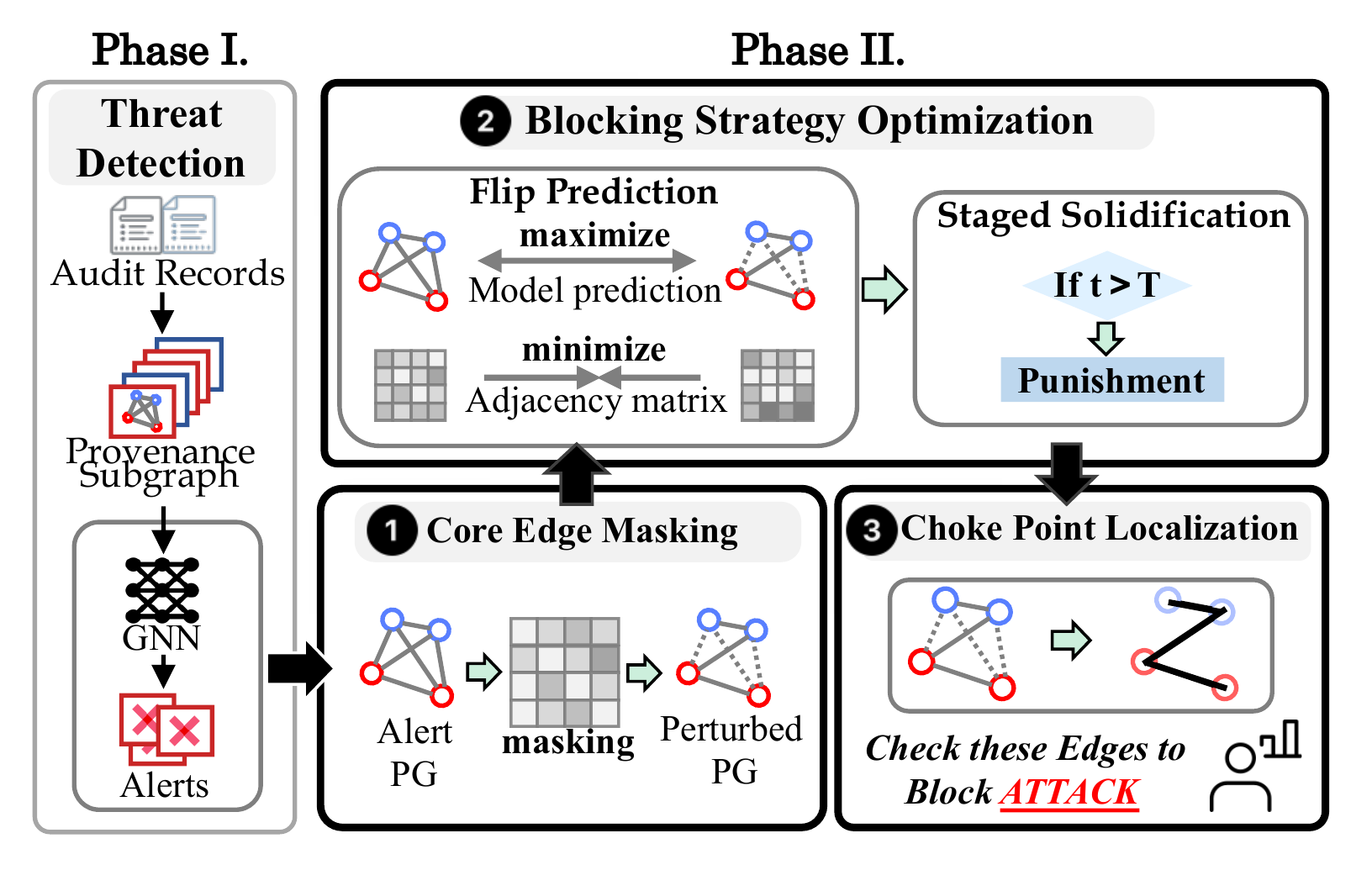}
	\vspace{-0.1in}
	\caption{\ProvX working mode with two phases.}
	% \Description{A figure illustrating four scenes.}
	\label{fig:ov}
	
\end{figure}

%
%As shown in Figure \ref{fig:ov}, \ProvX is a two-stage alert refinement framework designed to resolve alert flooding and enable attack blocking.
%
%
% \noindent\textbf{Phase I: Threat Detection (\ref{phase1})}: \ProvX uses a GNN-based detector to analyze numerous provenance subgraphs and output alerts, which are provenance subgraphs that encompass suspicious behaviors.
%
% \noindent\textbf{Phase II: Attack Mitigation (\ref{phase2})}: \ProvX refines all alerts, using a counterfactual explanation logic approach to output an actionable core edge list to block attack process. This phase consists of three steps:
%
% First, \ding{182} \textbf{Core Edge Masking} transforms the discrete edge perturbation search problem in alert refinement into a continuous core edge mask learning task.
% Then, \ding{183} \textbf{Blocking Strategy Optimization} is guided by a dual-objective optimization framework that ensures both alert neutralization and minimal blocking cost, supplemented by a staged solidification strategy to enhance stability.
% Finally, \ \  \ding{184}\textbf{Choke Point Localization} generating an immediately actionable blocking solution.

As shown in Figure \ref{fig:ov}, \ProvX is a two-phase alert refinement framework designed to resolve alert flooding and support attack blocking. 
%The core input of \ProvX is an alert-associated provenance subgraph: a suspicious region of host activity that has been raised by a detector or assembled during SOC investigation.

\noindent\textbf{Phase I: Threat Detection (\ref{phase1})}. This phase obtains suspicious provenance subgraphs for downstream refinement. In this paper, we instantiate it with a GNN-based graph-level detector over provenance subgraphs. In deployment, this acquisition step can also be replaced by mature SOC alert-discovery pipelines, such as existing PIDS alerts, EDR/SIEM alarms expanded through provenance queries, network indicators mapped to host events, or analyst-guided investigation.

\noindent\textbf{Phase II: Attack Mitigation (\ref{phase2})}. \ProvX refines each alerted subgraph with counterfactual reasoning and outputs a core edge list as blocking-oriented response guidance. This phase consists of three steps: First, \ding{182} \textbf{Core Edge Masking} relaxes the discrete edge intervention search into a continuous core edge mask learning task. Then, \ding{183} \textbf{Blocking Strategy Optimization} uses a dual-objective optimization framework to flip the detector's prediction while minimizing the simulated intervention cost, supplemented by staged solidification to improve stability. Finally, \ding{184} \textbf{Choke Point Localization} converts the learned mask into a concise core edge list for analyst review and possible host-level enforcement.

%Core Edge Masking: Transforms the discrete edge perturbation search problem in alert refinement into a continuous Core Edge Mask learning task.
%
%Refinement Optimization Objective: Is guided by a dual-objective optimization framework that ensures both alert neutralization and minimal blocking cost, supplemented by a Staged Solidification strategy to enhance stability.
%
%Generating Actionable Edge List: Outputs the final core edge list as an immediately actionable blocking solution.

\subsection{Threat Detection}
\label{phase1}

%This phase is based on an efficient APT detection model. Given host audit log data, it is constructed into a provenance graph containing system entities and their interaction information. The goal of this phase is to utilize a GNN to analyze this graph and generate alerts if it represents APT activity. This process aims to identify key patterns or substructures within the provenance graph that are related to APT behavior.
%
%Existing detectors use GNNs for autoencoding \cite{jia2024magic}, semi-supervised detection \cite{qiao2024slot}, and others. However, unlike the detection task, the core of \ProvX more lies in alert refinement. Therefore, we simply map APT detection to a GNN-based subgraph classification task to obtain alerts. Specifically, we focus on particular structures within provenance subgraphs (e.g., suspicious processes, abnormally accessed files, or interactions between critical system events) and attempt to classify these structures and their local environments to determine if they are involved in malicious activities.
%\noindent Next, we detail the steps of provenance detection:

Phase I is an instantiation of how \ProvX obtains alert-associated provenance subgraphs. In real SOC deployments, suspicious subgraphs may come from existing PIDS, endpoint alarms, network indicators, or analyst-guided provenance expansion. We construct this phase using a GNN-based graph-level detector and a Louvain-based partitioning procedure to approximate the alert-subgraph acquisition process.

Existing detectors use GNNs for autoencoding \cite{jia2024magic}, semi-supervised detection \cite{qiao2024slot}, and other detection objectives. However, the main contribution of \ProvX lies in downstream alert refinement rather than in proposing a new detector. Therefore, we map APT detection to a GNN-based subgraph classification task and use it to obtain alerted provenance subgraphs for Phase II. Specifically, we focus on local structures within provenance subgraphs, such as suspicious processes, abnormally accessed files, or interactions between critical system events, and classify whether these structures and their local context are associated with malicious activity.
\noindent Next, we detail this experimental instantiation:

\noindent\textbf{(1) Provenance Graph Construction}. Host audit logs are first processed and converted into a provenance graph $G_{host} =(V,E)$, where nodes $V$ represent system entities or events that occur, and edges $E$ represent causal dependencies between these entities (e.g., a process creating a file, a process sending data over the network, etc.). This provenance graph $G_{host}$ serves as the basis for subsequent subgraph extraction.

\noindent\textbf{(2) Subgraph Extraction}. We use a Louvain-based community partitioning procedure~\cite{blondel2008fast} to extract provenance subgraphs with manageable sizes. Benign subgraphs consist entirely of benign system entities, while malicious subgraphs contain at least one attack-related entity or attack stage. To preserve contextual completeness, we allow important hub nodes to appear in multiple extracted subgraphs when needed. This partitioning procedure is used to create evaluation subgraphs with controllable size and contextual completeness; it is not required when a deployment already provides alert-associated provenance subgraphs through mature SOC workflows. Alternative ways to construct alert-associated subgraphs are discussed in Section \ref{subgraph}.

We extract $N$ provenance subgraphs $\{G_1, G_2, \dots, G_N\}$, where each subgraph $G_k$ corresponds to a ground truth label $Y_k \in \{0, 1\}$. Here, 0 indicates that all system entities within the subgraph are benign (i.e., the subgraph is benign), while 1 indicates the presence of malicious system entities.

\noindent\textbf{(3) Threat Detection}.
The goal of provenance detection is to predict a label for each extracted subgraph $G_k$, and trigger an alert when it is associated with an APT. By inputting the provenance subgraphs into GNN models, we learn a mapping function $f(\cdot)$ to get the predicted probability $P(c | G_k)$. The final result label $\hat{Y}_K$ for $G_k$ is determined by chosing the highest probability:
\begin{equation}
	\hat{Y}_k = \operatorname*{arg\,max}_{c \in \{0,1\}} P(c | G_k).
\end{equation}
%For model selection, we adopt standard GNN model frameworks widely used in the industry (including GCN, GAT and GraphSAGE)\footnote{It is worth noting that most of the custom models in these works\cite{rehman2024flash, wang2022threatrace, jia2024magic, qiao2024slot, goyal2024r, zengy2022shadewatcher} are also based on these basic GNN architectures.}. 

For model selection, we adopt standard differentiable GNN frameworks, including GCN, GAT, and GraphSAGE\footnote{It is worth noting that many custom models in prior PIDS works\cite{rehman2024flash, wang2022threatrace, jia2024magic, qiao2024slot, goyal2024r, zengy2022shadewatcher} are also built on these basic GNN architectures.}. While existing systems such as \Slot\cite{qiao2024slot} may include non-differentiable components such as reinforcement learning, the current counterfactual optimization in \ProvX requires a compatible differentiable detector to compute prediction flips. Thus, the alert-discovery source is replaceable, but the Phase II optimization currently assumes access to a differentiable GNN-based detector.

\subsection{Attack Mitigation}
\label{phase2}

% \begin{figure}
% 	\centering
% 	\includesvg[width=0.5\textwidth]{figure/ov5_29.svg}
% 	\caption{The framework of \ProvX.}
% 	\label{fig:ov}
% 	\vspace{-0.1in}
% \end{figure}

%In this section, we propose \ProvX, Fig \ref{fig:ov} shows \ProvX's working mode. \ProvX is a counterfactual explanation framework for provenance graphs, which comprises the following three components:
%
%\noindent\ding{182} \textbf{Provenance Subgraph Perturbation(\ref{perturbation})}. To perform structural changes on the provenance subgraph, we represent perturbations to the provenance graph by learning adjustable edge masks, thereby transforming the discrete graph edit search problem for counterfactual changes into a continuous mask learning task.
%
%\noindent\ding{183} \textbf{Counterfactual Reasoning(\ref{cfx})}. We seek a minimal graph perturbation scheme that can explicitly flip the security model's original judgment of an APT threat. \ProvX constructs a dual optimization objective framework comprising a threat misclassification guidance loss and an attack path minimal perturbation loss, and introduces a staged solidification strategy to focus on core attack elements.
%
%\noindent\ding{184} \textbf{Generate Explanations(\ref{generate})}. After the above optimization learning process, we sort the final edge masks according to their importance and transform them into independent counterfactual subgraphs to obtain a complete explanation.

\subsubsection{Core Edge Masking}
\label{perturbation}

The first step of the \ProvX attack mitigation algorithm is to simulate edge-level interventions within an alert-associated provenance subgraph $G_k$. Since APT alerts often depend on structural dependencies, our objective is to learn a core edge mask that can flip the detector's alert at minimal simulated intervention cost. We focus on perturbing system entity interactions. For an alert subgraph $G_k$, its structure is represented by the adjacency matrix $A_k \in \{0, 1\}^{n\times n}$.

Directly searching for a discrete, binary ($\{0, 1\}$) blocking mask is an NP-hard combinatorial optimization problem that cannot be solved directly. Therefore, \ProvX relaxes this discrete search problem into a continuous, differentiable optimization task. Specifically, we introduce a learnable, continuous-valued parameter matrix $\hat{M}_k \in \mathbb{R}^{n \times n}$ with the same dimensions as $A_k$. We then use the sigmoid function $\sigma(\cdot)$ to map this matrix into the (0, 1) range, generating a soft mask. This soft mask represents the probability of retaining each edge. The perturbed adjacency matrix $\tilde{A}_k$ is then defined as:
\begin{equation}\tilde{A}_k = A_k \odot \sigma(\hat{M}_k),
\end{equation}
where $\odot$ denotes element-wise multiplication. The $\sigma(\cdot)$ function provides a smooth transition between "retaining" (a value near 1) and "blocking" (a value near 0). Consequently, starting from a randomly initialized mask, $\hat{M}_k$ can be optimized via gradient descent to learn the minimal set of edges required to flip the detector's alert.

\subsubsection{\textbf{Blocking Strategy Optimization}}
\label{cfx}
We conduct the core optimization task by learning masks for the provenance subgraphs (i.e., alerts) predicted by the security model as malicious. The core idea of algorithm is to determine the minimum perturbation of the provenance subgraph (i.e., the minimal blocking set), which is achieved by establishing an optimization task and optimizing the composite loss function. It is worth mentioning that our blocking strategy optimization task only focuses on the learning of edge masks and does not involve the learning of the GNN security model. The parameters in the GNN security model are fixed. The specific workflow is as follows: We perturb the target alert subgraph as described in section \ref{perturbation} to generate a perturbed subgraph $\tilde{G}_k$, and input it together with the original subgraph $G_k$ into the well-trained GNN-based security model we built in Phase I to generate the corresponding prediction results, which are:
\begin{equation}   \hat{Y}_k = \text{GNN}(A_k), \  \    \tilde{Y}_k = \text{GNN}(\tilde{A}_k).
\end{equation}
We learn the edge mask $\hat{M}_k$ based on the optimization objective of counterfactual reasoning to determine the minimal perturbation. The mask learning process is achieved by minimizing a composite loss function \textbf{$\mathcal{L}_{\textbf{\ProvX}}$} over $T$ training epochs. This loss function simulates the trade-off between attempting to flip the detector's alert (i.e., change the model's prediction) and minimizing the response cost (i.e., minimizing modifications to the original activity logs). It consists of a primary counterfactual loss $\mathcal{L}_{\textbf{CFX}}$ and a staged solidification loss $\mathcal{L}_{\textbf{S}}$.

\noindent \textbf{(1) Primary Counterfactual Loss $\mathcal{L}_{\textbf{CFX}}$.} The guiding principle of $\mathcal{L}_{\textbf{CFX}}$ is to find modifications to the provenance graph that can change the detector's malicious prediction while keeping the perturbation (i.e., blocking cost) as small as possible. Thus, $\mathcal{L}_{\textbf{CFX}}$ is made up of two loss terms, Prediction Flip Loss and Mask Distance Loss. The Prediction Flip Loss $\mathcal{L}_{\textbf{Pred}}$ drives the model to find a modification to the provenance graph such that the modified graph is no longer classified under the original threat category (i.e., the alert is cleared). We achieve this by minimizing the probability corresponding to the original threat prediction $\hat{Y}_k$, which encourages the model to significantly reduce its confidence in the original threat assessment:
\begin{equation}    \mathcal{L}_{\textbf{Pred}} = P(\hat{Y}_k, \tilde{A}_k).\end{equation}
The Mask Distance Loss $\mathcal{L}_{\textbf{dist}}$ aims to ensure that the identified provenance graph modification is as small as possible. This means identifying only those most critical, core activity segments or system interactions that form the root cause of the alert. If a masking scheme requires modifying a large number of event correlations to make the GNN model change its prediction, then this loss term will be larger:
\begin{equation}    \mathcal{L}_{\textbf{dist}} = \text{BinaryCrossEntropy}(\sigma(\hat{M}_k), A_k).
\end{equation}
Then, we integrate the above two loss terms into the primary counterfactual loss function to optimize them collaboratively:
\begin{equation}    \mathcal{L}_{\textbf{{CFX}}} = \alpha \cdot \mathcal{L}_{\textbf{pred}} + (1-\alpha) \cdot \mathcal{L}_{\textbf{dist}},
\end{equation}where $\alpha$ is responsible for regulating the trade-off between the predict flip loss term and the mask distance loss term. A higher $\alpha$ prioritizes flipping the detector's alert at the expense of larger perturbations, while a lower $\alpha$ focuses more on minimizing graph modifications.

\noindent \textbf{(2) Staged Solidification Loss $\mathcal{L}_{\textbf{S}}$.} To provide security analysts with clearer and more reliable localization results for critical attack paths, \ProvX introduces a staged solidification mechanism. This mechanism is controlled by a solidification factor $\gamma_S$, a solidification stage starting ratio $R_S \in [0,1]$, and confidence thresholds $\tau_{low}$ and $\tau_{high}$. Its core idea is to further reinforce the judgment on event correlations that have already been initially identified as extremely important or unimportant for prediction flipping in the later stages of explanation model training, making their contributions more prominent.

The solidification process is activated after an initial exploratory learning phase, specifically starting from the $T_{start}$-th epoch, and its strength is controlled by the solidification factor $\gamma_S$.
%\noindent \textbf{Mask Snapshot and Confident Edge Identification}. 
At the $T_{start}$ epoch, the currently activated edge mask (specifically for $\sigma(\hat{M}_k)$, denoted as $M^k_{snap} = \sigma(\hat{M}^{(T_{start})}_{k})$ is recorded. Then, based on predefined low confidence threshold $\tau_{low}$ (confident threshold low) and high confidence threshold $\tau_{high}$ (confident threshold high), two sets of edge indices are identified: 
\begin{itemize}
 \item For low confidence edge index set $I_{low} = \{i|m^k_{snap,i}<\tau_{low}\}$ (edges whose mask values are already close to 0 in the snapshot),
  \item For high confidence edge index set $I_{high} = \{j|m^k_{snap,j}>\tau_{high}\}$ (edges whose mask values are already close to 1 in the snapshot).
\end{itemize}

For epochs $t>T_{start}$, the solidification penalty $\mathcal{L}_S$ is calculated as follows:
\begin{equation}
	\begin{split}
		\mathcal{L}_S = & \sum_{i \in I_{low}} (\sigma(\hat{M}_{k,i}^{(t)}) - m_{snap,i}^k)^2 \\
		& + \sum_{j \in I_{high}}(\sigma(\hat{M}_{k,j}^{(t)}) - m_{snap,j}^k)^2.
	\end{split}
\end{equation}
This penalty term targets edges whose mask values already showed a clear trend. This encourages the mask values to consolidate towards 0 or 1. 

Finally, the total loss function $\mathcal{L}_{\textbf{\ProvX}}$ is the sum of the primary counterfactual loss and the weighted solidification loss:
\begin{equation}    \mathcal{L}_{\text{\ProvX}} = \mathcal{L}_{\text{CFX}} + \gamma_S \cdot \mathcal{L}_S,
\end{equation}
where the $\gamma_S$ parameter is used to fix the weight of the solidification loss.

\subsection{Choke Point Localization}
\label{generate}

In this phase, we transform the optimization results from the blocking strategy algorithm (\ref{cfx}) into a core edge list that security analysts can prioritize for review and possible enforcement.

After the optimization process is complete, we obtain a final continuous event mask, $\sigma(\hat{M}^*_k)$. By calculating $(1 - \sigma(\hat{M}^*_k))$, we derive a blocking priority score for each edge in the alert graph. A high score indicates that the edge is highly necessary for the detector's malicious prediction under the learned counterfactual objective. Therefore, high-scoring edges represent core behavioral links that analysts should prioritize for review and candidate blocking. This score reflects model-level decision necessity rather than the operational ease or safety of enforcing a host-level action.

We select the top-$K$ edges with the highest blocking priority score. In our framework, $K$ is a user-definable response budget. This budget allows analysts, based on their real-world operational pressure, to trade off between response cost and the completeness of alert refinement. For example, an analyst facing alert fatigue can set a low $K$ to obtain the most compact list of candidate choke points. Our objective is to flip the detector's alert with the smallest simulated intervention, thereby producing a concise edge list for downstream review.

Finally, we extract the subgraph $\tilde{G}^*_k$ formed by these top-$K$ core edges. This subgraph is the blocking-oriented core edge list output by \ProvX. For model-level evaluation, we instantiate the simulated intervention by removing the selected edges and suppressing the features of their incident endpoint nodes. This joint structural-attribute intervention reflects both the disruption of selected event correlations and the removal of their local behavioral evidence from the detector's input. Security analysts can interpret the result as:
\textit{"Under the detector's current decision logic, this alert is most sensitive to the following $K$ critical edges; these edges should be prioritized for review and possible blocking."}

\section{\textbf{Detection-Mitigation Feedback Loop}}
\label{feedback}

\begin{figure}[t]
	\centering
	\includegraphics[width=0.47\textwidth]{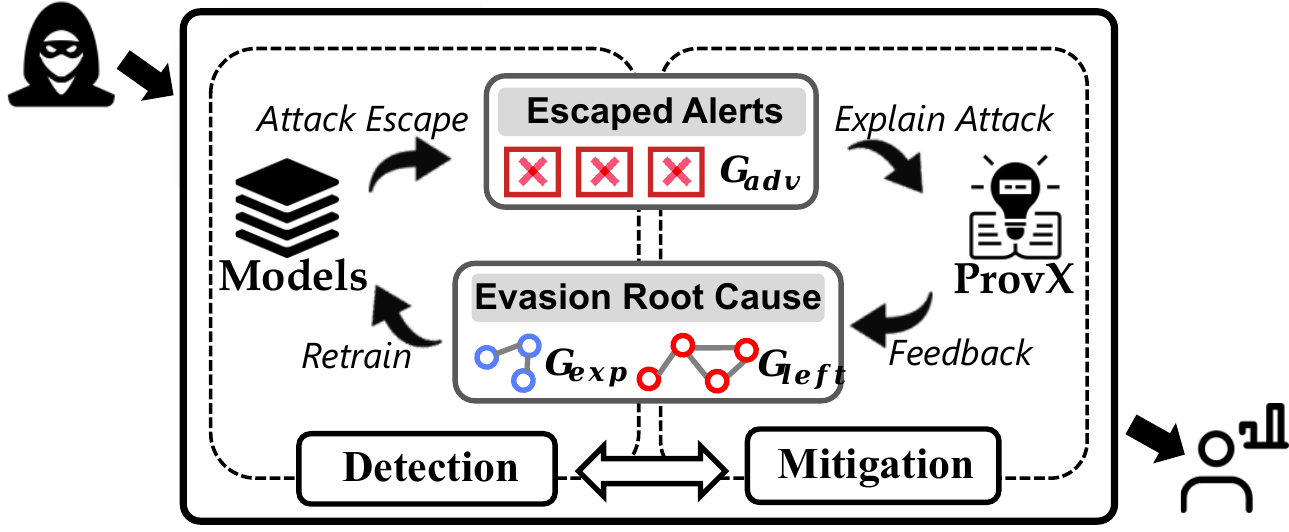}
	%\vspace{-0.1in}
	\caption{Mitigation-driven feedback loop for detection enhancement.
	}
	% \Description{A figure illustrating four scenes.}
	\label{fig:feedback}
	
\end{figure}

Although \ProvX primarily analyzes malicious alerts for blocking-oriented refinement, the same counterfactual logic can also be used in reverse for false-negative analysis. When a malicious activity deceives a PIDS through camouflage or obfuscation\cite{goyal2023sometimes} and is misclassified as benign, we refer to it as an adversarial evasion. In this setting, the question is no longer ``which structures make the detector predict malicious?'', but rather ``which added or surrounding structures make the detector maintain a benign prediction for a known malicious behavior?''

We therefore explore a preliminary feedback framework guided by security analysts. Let $G_m$ denote an original malicious provenance subgraph. Following the mimicry attack setting in \cite{goyal2023sometimes}, we construct an adversarial subgraph $G_{adv}$ by injecting benign-looking provenance structures into $G_m$ and retain the samples that evade the detector. For such a false-negative sample, \ProvX is assigned a reversed objective: it searches for a minimal structure $G_{exp}$ whose removal flips the detector's prediction from benign back to malicious. We interpret $G_{exp}$ as candidate evasion-associated context, rather than as definitive proof of attacker intent.

After analyst verification, we remove the localized edges in $G_{exp}$ from $G_{adv}$ and denote the remaining residual graph as $G_{left}$. We then use both $G_{adv}$ and $G_{left}$ as malicious training examples. These two samples play different roles. $G_{adv}$ teaches the detector the complete obfuscated attack pattern, while $G_{left}$ preserves the residual malicious behavior after removing the localized evasion-associated context. This paired feedback is intended to improve robustness against similar false negatives while keeping analyst validation in the loop.

Analyst verification is important because the localized $G_{exp}$ may contain ambiguous benign context or low-confidence structures rather than a definitive evasion cause. We therefore do not directly treat every localized structure as an automatic training signal. Instead, analyst verification filters unreliable feedback and reduces the risk of increasing false positives during retraining. In addition, using the full adversarial graph $G_{adv}$ together with the residual graph $G_{left}$ avoids training only on a dismantled subgraph: $G_{adv}$ exposes the detector to the complete obfuscated attack, while $G_{left}$ reinforces the remaining attack core. This design aims to reduce false negatives against similar evasion patterns without unnecessarily weakening the detector's boundary between benign and malicious behavior.

We present this feedback loop as a preliminary extension of \ProvX, not as a fully general adversarial-training framework. Its effectiveness is evaluated in Section \ref{RQ6}.

\section{Evaluation}
\label{eval}

% Evaluation

% 在本节中，我们评估了\ProvX在解释攻击方面的有效性。我们的目标是回答以下研究问题：
% RQ1: 安全模型对恶意溯源子图的检测能力如何？
% RQ2: 与SOTA解释器相比较，\ProvX反事实解释的性能如何？
% RQ3: \ProvX关键组件的有效性如何？
% RQ4: 超参数对\ProvX解释性能的影响？
% RQ5: \ProvX的overhead如何？
% RQ7: \ProvX对于对抗性攻击的解释鲁棒性如何？

In this section, we evaluate the effectiveness of \ProvX. Our goal is to answer the following research questions:
\begin{itemize} 

	\item \textbf{RQ1}. \textbf{Localization Ability:} How accurately does \ProvX's core edge list hit real-world attack steps? 
	\item \textbf{RQ2}. \textbf{Blocking Effectiveness:} What is the blocking capability of the core edges localized by \ProvX?
	\item \textbf{RQ3}. \textbf{SecOps Cost-Benefit Analysis:} How does \ProvX's \textit{Cost} vs. \textit{Benefit} trade-off compare to SOTA PIDS? 
	\item \textbf{RQ4}. \textbf{Hyperparameter Sensitivity:} What is the impact of key hyperparameters on \ProvX's mitigation performance?
	\item \textbf{RQ5}. \textbf{System Overhead:} What is the computational overhead of the \ProvX framework? 
	\item \textbf{RQ6}. \textbf{Adversarial Analysis:} How can \ProvX be used to analyze and defend against adversarial evasion attacks? 
\end{itemize}

\subsection{Experimental Setup}

\subsubsection{Datasets} Consistent with existing APT detection works, we conduct our experimental evaluation on two widely used APT datasets: the DARPA TC (Transparent Computing program) and the DARPA OpTC (Operationally Transparent Cyber)\footnote{Existing PIDS based on provenance graphs all use one\cite{zengy2022shadewatcher} or more\cite{wang2022threatrace, cheng2024kairos} of these datasets to verify APT detection performance. Our work refers to \cite{jia2024magic,rehman2024flash,qiao2024slot} and selects the above datasets.}.
The DARPA TC dataset\cite{DARPA} was collected during adversarial engagements in a realistic environment. In these scenarios, a red team employed various exploits and attack techniques to compromise target hosts and obtain sensitive information, while a blue team worked to detect attacks and collect data on suspicious behavior through methods like host-level auditing and causal analysis. DARPA TC targets specific hosts for attack, and for our research, just like \cite{jia2024magic}, we select the \textbf{Cadets}, \textbf{Theia} and \textbf{Trace} scenarios. The DARPA OpTC dataset\cite{DARPA_OpTC} provides a broad view of benign and malicious audit records from approximately 1,000 hosts.
To improve training quality, we perform random negative sampling on the scarce malicious behavior subgraphs to obtain a balanced dataset. We split the data into training, validation, and test sets at a ratio of \textit{7:1:2}. 

\begin{table}[h]

	\resizebox{0.48\textwidth}{!}{
		\begin{threeparttable}
			\begin{tabular}{@{}c|cccccc@{}}
				\toprule
				\textbf{GNN type}  & \textbf{Acc}    & \textbf{Pr}     & \textbf{Rec}    & \textbf{F$_1$}     & \textbf{AUC}    & \textbf{FPR\tnote{1}}    \\ \midrule
				GCN       & 0.7379 & 0.5757 & 0.6340 & 0.6034 & 0.8006 & 0.2144  \\ \midrule
				GAT       & 0.9126 & 0.8464 & 0.8824 & 0.8640 & 0.9670 & 0.0735  \\ \midrule
				GraphSAGE & 0.9404 & 0.9276 & 0.8791 & 0.9027 & 0.9784 & 0.0315  \\ \bottomrule
			\end{tabular}
			\begin{tablenotes}
				\item[1] FPR (False Positive Rate), a detection system with a lower false positive rate can better avoid false positive during detection.
				% \item[2] TPR (True Positive Rate) represents the ratio of all real attacks that the system can successfully detect. A high TPR means that the system can effectively identify most real attacks.
			\end{tablenotes}
		\end{threeparttable}
	}
		\caption{DARPA OpTC detection performance on three GNNs.}
\label{detection_performance_optc}
\end{table}

\begin{table}[h]

	\resizebox{0.48\textwidth}{!}{
		\begin{threeparttable}
			\begin{tabular}{@{}c|c|cccccc@{}}
				\toprule
				\textbf{GNN}                                                 & \textbf{Datasets} & \textbf{Acc}   & \textbf{Pr}    & \textbf{Rec}   & \textbf{F$_1$}   & \textbf{AUC}   & \textbf{FPR}   \\ \midrule
				\multirow{3}{*}{GCN}                                         & Cadets            & 0.9972         & 0.9999         & 0.9286         & 0.9630         & 0.9938         & 0.0001         \\
				& Theia             & 0.9612         & 0.8491         & 0.7031         & 0.7692         & 0.8947         & 0.0127         \\
				& Trace             & 0.9968         & 0.9778         & 0.9635         & 0.9706         & 0.9915         & 0.0013         \\ \midrule
				\multirow{3}{*}{GAT}                                         & Cadets            & 0.9944         & 1.0000         & 0.8571         & 0.9231         & 0.9647         & 0.0001         \\
				& Theia             & 0.9554         & 0.7705         & 0.7344         & 0.7520         & 0.9209         & 0.0222         \\
				& Trace             & 0.9968         & 0.9850         & 0.9562         & 0.9704         & 0.9918         & 0.0009         \\ \midrule
				\multirow{3}{*}{\begin{tabular}[c]{@{}c@{}}Graph\\ SAGE\end{tabular}} & Cadets            & 0.9986         & 0.9999         & 0.9643         & 0.9818         & 0.9931         & 0.0001         \\
				& Theia             & 0.9885         & 0.9828         & 0.8906         & 0.9344         & 0.9617         & 0.0016         \\
				& Trace             & 0.9955         & 0.9701         & 0.9489         & 0.9594         & 0.9907         & 0.0017         \\ \bottomrule
			\end{tabular}
			% \begin{tablenotes}
				% 	% \item[1] FPR (False Positive Rate), a detection system with a lower false positive rate can better avoid false positive attacks during detection.
				%     % \item[2] TPR (True Positive Rate) represents the ratio of all real attacks that the system can successfully detect. A high TPR means that the system can effectively identify most real attacks.
				% \end{tablenotes}
		\end{threeparttable}
	}
		\caption{DARPA E3 detection performance on three GNNs.}
\label{detection_performance_e3}
\end{table}

\subsubsection{Implementation Details} 

We detail the implementation of our framework's two main components: \noindent \textbf{Phase I: Threat Detection}. We implemented subgraph-level detection models using GCN, GAT, and GraphSAGE. Each model uses a 2-layer GNN architecture with ReLU/Dropout, followed by an average pooling layer and a 2-layer MLP classifier. These detectors were trained for 50 epochs using an Adam optimizer with a 0.001 learning rate. As shown in Tables \ref{detection_performance_optc} and \ref{detection_performance_e3}, the models achieve high performance, providing a solid and reliable foundation for our Phase II task. 
\noindent \textbf{Phase II: Attack Mitigation}. Each triggered alert subgraph is processed independently. We train the core edge mask for 200 epochs with a learning rate of 0.01. Key hyperparameters for the Staged Solidification strategy were set as follows: solidification ratio $R_S=0.6$, factor $\gamma_S=0.5$, and confidence bounds $\tau_{low}=0.05$ and $\tau_{high}=0.95$. A detailed analysis of the trade-off parameter $\alpha$ and the response budget $K$ is presented in Section \ref{RQ4}.

All experiments were performed on a server running Ubuntu 22.04.5 LTS with an Intel(R) Xeon(R) Silver 4208 CPU (14 cores, 14 threads, base frequency 2.10 GHz) and 256 GB of RAM.

\subsubsection{Baselines for Comparison} Our evaluation is designed to validate \ProvX's performance in the novel task of attack mitigation. This requires a two-pronged comparison, as no single baseline fully addresses this challenge.
First, to validate the algorithmic superiority of our localization ability(\textbf{RQ1}), we compare our counterfactual-based algorithm against established triage tools and SOTA explainable artificial intelligence (XAI) explainers. Second, to validate the mitigation ability(\textbf{RQ2}), we compare with XAI explainers. These XAI explainers are fact-based, including perturbation-based, decomposition-based, and gradient-based methods \cite{ying2019gnnexplainer, yuan2021explainability, luo2020parameterized, schnake2021higher, shrikumar2017learning, selvaraju2017grad}. Unlike these methods, which primarily seek subgraphs that preserve or explain the original prediction, \ProvX searches for a minimal intervention that changes the detector's prediction and ranks edges by blocking-oriented necessity.
And, to demonstrate our framework's operational value and SecOps advantage (\textbf{RQ3}), we compare our cost-benefit output against raw alerts produced by SOTA PIDS detectors. This group includes prominent systems such as \MAGIC\cite{jia2024magic}, \Slot \cite{qiao2024slot}, \FLASH \cite{rehman2024flash}, \ShadeWatcher \cite{zengy2022shadewatcher}, and \ThreaTrace \cite{wang2022threatrace}.

\begin{table}[t]

	\centering
	\resizebox{0.44\textwidth}{!}{
		\renewcommand{\arraystretch}{0.95}
		\begin{threeparttable}
			
			\begin{tabular}{@{}c|c|c|ccc@{}}
				\toprule
				\multirow{2}{*}{\begin{tabular}[c]{@{}c@{}}\textbf{GNN}\\ \textbf{Type}\end{tabular}} & \multirow{2}{*}{\textbf{Metrics\tnote{1}}} & \multirow{2}{*}{\begin{tabular}[c]{@{}c@{}}\textbf{DARPA}\\ \textbf{OpTC}\end{tabular}} & \multicolumn{3}{c}{\textbf{DARPA TC E3}} \\ \cmidrule(l){4-6} 
				&  &  & Cadets & Theia & Trace \\ \midrule
				\multirow{4}{*}{GCN}
				& Acc & 0.8402 & 0.9643 & 0.9375 & 1.0000 \\
				& Pr & 0.1009 & 0.9024 & 0.8934 & 0.9747 \\
				& Rec & 0.7573 & 0.1470 & 0.1182 & 0.1226 \\
				& F1 & 0.1683 & 0.2517 & 0.1970 & 0.1982 \\ \midrule
				\multirow{4}{*}{GAT}
				& Acc & 0.9148 & 0.9643 & 0.9688 & 1.0000 \\
				& Pr & 0.1293 & 0.9125 & 0.9033 & 1.0000 \\
				& Rec & 0.8169 & 0.2249 & 0.2244 & 0.2758 \\
				& F1 & 0.2118 & 0.3101 & 0.3342 & 0.3524 \\ \midrule
				\multirow{4}{*}{GraphSAGE}
				& Acc & 0.9033 & 1.0000 & 1.0000 & 1.0000 \\
				& Pr & 0.1306 & 0.9798 & 0.9431 & 0.9511 \\
				& Rec & 0.7035 & 0.3068 & 0.2914 & 0.3318 \\
				& F1 & 0.2006 & 0.4531 & 0.3826 & 0.4773 \\ \bottomrule
			\end{tabular}
			\begin{tablenotes}
				\item[1] \textbf{\ProvX} uses different security model, so the number of malicious subgraphs captured is different.
			\end{tablenotes}
		\end{threeparttable}
	}
		\caption{Localization Ability of \textbf{\ProvX} in DARPA OpTC(with $K$=10) and TC-E3(with $K$=30) on GCN, GAT, and GraphSAGE.}
\label{mingzhong_all}
\end{table}

\begin{table}[t]
	
	\resizebox{0.485\textwidth}{!}{
		\renewcommand{\arraystretch}{1.3}
		\begin{threeparttable}
			\setlength{\tabcolsep}{3pt}
			
			\begin{tabular}{@{}c|c|cccc@{}}
				\toprule
				\textbf{GNNs} & \textbf{Explainers\tnote{1}} & \textbf{Acc} & \textbf{Pr} & \textbf{Rec} & \textbf{F1} \\ \midrule
				\multicolumn{1}{c|}{} & \multicolumn{1}{c|}{GradCam} & $0.2165_{\dec{74.2}}$ & $0.0944_{\dec{6.4}}$ & $0.2473_{\dec{67.3}}$ & $0.1418_{\dec{15.7}}$ \\
				\multicolumn{1}{c|}{} & \multicolumn{1}{c|}{DeepLIFT} & $0.2639_{\dec{68.6}}$ & $0.0355_{\dec{64.8}}$ & $0.1832_{\dec{75.8}}$ & $0.0730_{\dec{56.6}}$ \\
				\multicolumn{1}{c|}{} & \multicolumn{1}{c|}{GNN-LRP} & $0.2526_{\dec{69.9}}$ & $0.0891_{\dec{11.7}}$ & $0.1774_{\dec{76.6}}$ & $0.1186_{\dec{29.5}}$ \\
				\multicolumn{1}{c|}{} & \multicolumn{1}{c|}{PGExplainer} & $\underline{0.7938}_{\dec{5.5}}$ & $0.0978_{\dec{3.1}}$ & $0.7147_{\dec{5.6}}$ & $0.1426_{\dec{15.3}}$ \\
				\multicolumn{1}{c|}{} & \multicolumn{1}{c|}{SubgraphX} & $0.7783_{\dec{7.4}}$ & $\underline{0.1019}_{\inc{1.0}}$ & $0.6413_{\dec{15.3}}$ & $0.1011_{\dec{39.9}}$ \\
				\multicolumn{1}{c|}{} & \multicolumn{1}{c|}{GNNExplainer} & $0.8402_{0.0\%}$ & $\textbf{0.1025}_{\inc{1.6}}$ & $\underline{0.7492}_{\dec{1.1}}$ & $\textbf{0.1698}_{\inc{0.9}}$ \\
				\multicolumn{1}{c|}{\multirow{-7}{*}{GCN}} & \multicolumn{1}{c|}{\cellcolor[HTML]{D9D9D9}\textbf{\ProvX} } & \cellcolor[HTML]{D9D9D9}$\textbf{0.8402}_{\inc{60.3}}$\tnote{2} & \cellcolor[HTML]{D9D9D9}$0.1009_{\inc{16.2}}$ & \cellcolor[HTML]{D9D9D9}$\textbf{0.7573}_{\inc{67.5}}$ & \cellcolor[HTML]{D9D9D9}$\underline{0.1683}_{\inc{35.2}}$ \\ \midrule
				\multicolumn{1}{c|}{} & \multicolumn{1}{c|}{GradCam} & $0.6570_{\dec{28.2}}$ & $0.1173_{\dec{9.0}}$ & $0.6867_{\dec{15.9}}$ & $0.1894_{\dec{10.6}}$ \\
				\multicolumn{1}{c|}{} & \multicolumn{1}{c|}{DeepLIFT} & $0.7956_{\dec{13.0}}$ & $0.1044_{\dec{19.3}}$ & $0.7039_{\dec{13.5}}$ & $0.1948_{\dec{8.0}}$ \\
				\multicolumn{1}{c|}{} & \multicolumn{1}{c|}{GNN-LRP} & $0.8540_{\dec{6.6}}$ & $0.1151_{\dec{10.7}}$ & $0.7666_{\dec{6.2}}$ & $0.1974_{\dec{6.8}}$ \\
				\multicolumn{1}{c|}{} & \multicolumn{1}{c|}{PGExplainer} & $0.8905_{\dec{1.6}}$ & $\underline{0.1188}_{\dec{8.1}}$ & $\underline{0.8007}_{\dec{1.9}}$ & $\underline{0.2021}_{\dec{5.1}}$ \\
				\multicolumn{1}{c|}{} & \multicolumn{1}{c|}{SubgraphX} & $\underline{0.9015}_{\dec{1.4}}$ & $0.1162_{\dec{10.6}}$ & $0.7938_{\dec{3.6}}$ & $0.1950_{\dec{8.0}}$ \\
				\multicolumn{1}{c|}{} & \multicolumn{1}{c|}{GNNExplainer} & $0.8759_{\dec{4.2}}$ & $0.1130_{\dec{12.1}}$ & $0.7848_{\dec{2.2}}$ & $0.1977_{\dec{6.7}}$ \\
				\multicolumn{1}{c|}{\multirow{-7}{*}{GAT}} & \multicolumn{1}{c|}{\cellcolor[HTML]{D9D9D9}\textbf{\ProvX} } & \cellcolor[HTML]{D9D9D9}$\textbf{0.9148}_{\inc{10.3}}$ & \cellcolor[HTML]{D9D9D9}$\textbf{0.1293}_{\inc{13.3}}$ & \cellcolor[HTML]{D9D9D9}$\textbf{0.8169}_{\inc{8.0}}$ & \cellcolor[HTML]{D9D9D9}$\textbf{0.2118}_{\inc{8.0}}$\\ \midrule
				\multicolumn{1}{c|}{} & \multicolumn{1}{c|}{GradCam} & $0.7434_{\dec{17.7}}$ & $0.0784_{\dec{39.8}}$ & $0.6202_{\dec{11.9}}$ & $0.1311_{\dec{34.6}}$ \\
				\multicolumn{1}{c|}{} & \multicolumn{1}{c|}{DeepLIFT} & $0.7199_{\dec{20.2}}$ & $0.0727_{\dec{43.9}}$ & $0.5925_{\dec{15.8}}$ & $0.1230_{\dec{38.7}}$ \\
				\multicolumn{1}{c|}{} & \multicolumn{1}{c|}{GNN-LRP} & $0.6841_{\dec{24.3}}$ & $0.0877_{\dec{32.8}}$ & $0.5800_{\dec{17.6}}$ & $0.1206_{\dec{40.0}}$ \\
				\multicolumn{1}{c|}{} & \multicolumn{1}{c|}{PGExplainer} & $0.8922_{\dec{1.2}}$ & $0.1233_{\dec{5.6}}$ & $0.6748_{\dec{4.1}}$ & $0.1779_{\dec{11.3}}$ \\
				\multicolumn{1}{c|}{} & \multicolumn{1}{c|}{SubgraphX} & $0.9033_{0.0\%}$ & $\underline{0.1298}_{\dec{0.6}}$ & $\textbf{0.7124}_{\inc{1.3}}$ & $\textbf{0.2147}_{\inc{7.0}}$ \\
				\multicolumn{1}{c|}{} & \multicolumn{1}{c|}{GNNExplainer} & $\textbf{0.9108}_{\inc{0.8}}$ & $0.1247_{\dec{4.5}}$ & $0.7002_{\dec{0.5}}$ & $0.1994_{\dec{0.6}}$ \\
				\multicolumn{1}{c|}{\multirow{-7}{*}{\begin{tabular}[c]{@{}c@{}}Graph\\ SAGE\end{tabular}}} & \cellcolor[HTML]{D9D9D9}\textbf{\ProvX} & \cellcolor[HTML]{D9D9D9}$\underline{0.9033}_{\inc{11.7}}$ & \cellcolor[HTML]{D9D9D9}$\textbf{0.1306}_{\inc{27.1}}$ & \cellcolor[HTML]{D9D9D9}$\underline{0.7035}_{\inc{8.8}}$ & \cellcolor[HTML]{D9D9D9}$\underline{0.2006}_{\inc{24.5}}$ \\ \bottomrule
			\end{tabular}
			
			\begin{tablenotes}
				\item[-] \textbf{Bold} denotes the best results, and \underline{underline} denotes the second-best results.
				\item[] $\dec{}$ represents the percentage of reduction, $\inc{}$ represents the percentage of improvement.
				\item[1] We adopt the hyperparameter configuration of previous work\cite{hu2023interpreters} to set up the baselines explainers.
				\item[2] Improvement compared to the average of all detection systems except \ProvX.

			\end{tablenotes}
			
		\end{threeparttable}
	}
	\caption{Comparison of \textbf{\ProvX} and XAI Baselines’ performance in explaining attack correlation on three GNNs, in DARPA OpTC.}
\label{mingzhong_sota_comparison}
\end{table}

\begin{table}[t]
	\centering

	\resizebox{0.4\textwidth}{!}{
		\renewcommand{\arraystretch}{1.3}
		\begin{threeparttable}
		\begin{tabular}{@{}c|cccc@{}}
			\toprule
			\textbf{Method} & \textbf{Accuracy} & \textbf{Precision} & \textbf{Recall} & \textbf{F1-score} \\ \midrule
			NoDoze \cite{hassan2019nodoze} & 0.8929 & 0.8595 & 0.2661 & 0.3994 \\
			DepImpact \cite{fang2022back} & \underline{0.9643} & \underline{0.9472} & \textbf{0.3643} & 0.4926 \\ \midrule
			\cellcolor[HTML]{D9D9D9}\textbf{\ProvX (Ours)} & \textbf{1.0000} & \textbf{0.9798} & \underline{0.3068} & \textbf{0.4531} \\ \bottomrule
		\end{tabular}

	\begin{tablenotes}
		\small
		\item[-] Results for NoDoze and DepImpact are aligned to the same budget $K=30$ for structural comparison. \ProvX results are based on the GraphSAGE backbone.
	\end{tablenotes}
	\end{threeparttable}
	}
		\caption{Comparison of \textbf{\ProvX} against Security Triage Baselines (NoDoze and DepImpact) in DARPA TC-E3 Scenarios.}
\label{security_baseline_comparison}
\end{table}

\subsection{Localization Ability \textit{(RQ1)}}
\label{RQ1}

The core output of the \ProvX framework is an actionable core edge list. A critical operational question is: to what extent does this list hit the real-world attack steps? Therefore, our first research question evaluates the localization ability of \ProvX.
We use the collection of malicious labels from existing works~\cite{jia2024magic, wang2022threatrace, rehman2024flash} as the ground truth for evaluation. After \ProvX analyze an alert subgraph, we use Accuracy, Precision, Recall and F1-score as evaluation metrics for localization hit ability. 

Specifically, if an edge within the Top-$K$ core edge list provided by \ProvX corresponds to a real attack, we consider it a Hit. Accuracy is defined as the proportion of malicious alert graphs in the test set for which \ProvX achieves at least one such hit. Precision verifies what proportion of the key events found by \ProvX are real key attack steps. Recall verifies what proportion of all real key attack steps are successfully found by \ProvX. The F1-score is computed per alert subgraph and then averaged across alert subgraphs.

Table \ref{mingzhong_all} presents the evaluation results of \ProvX across all datasets and GNN models, while Table \ref{mingzhong_sota_comparison} shows the comparison between \ProvX and the baseline localization algorithms. The results show that \ProvX achieves high Accuracy across all datasets, with an average Accuracy of 0.8861, demonstrating that at least a portion of the edges considered important by \ProvX overlaps with real attacks. We found that the prevalence of star structures (such as those related to \texttt{Nginx}) in the DARPA TC dataset significantly affects the learning trend of the detection model. The $K$ key edges found by \ProvX under the given response budget cannot fully cover the numerous interactions of the hub nodes. The attack distribution in OpTC is more scattered and complex; \ProvX can obtain multiple important links of the scattered attack links, but these important edges may be crucial to triggering the alert (model prediction), rather than the real steps annotated by experts.

In the comparative experiments (Table \ref{mingzhong_sota_comparison}), \ProvX outperforms the baseline algorithms in the vast majority of cases, demonstrating the effectiveness of counterfactual logic in the APT alert refinement domain. Among the baselines, the perturbation-based methods (GNNExplainer, PGExplainer, SubgraphX) perform slightly better than the decomposition-based or gradient-based methods, but \ProvX's localization accuracy remains consistently ahead. The main reason is the difference in optimization objective. Fact-based explainers tend to identify edges that are salient or correlated with the existing malicious prediction, which can include benign but highly connected background structures in noisy provenance graphs. In contrast, \ProvX explicitly optimizes for a minimal intervention that flips the detector's prediction, making the selected edges more aligned with model-level necessity. In addition, staged solidification stabilizes mask learning by locking high-confidence retain/remove trends in later epochs, which reduces oscillation around high-degree provenance hubs.

Compared to established security triage baselines(Table \ref{security_baseline_comparison}) and under an equivalent alert reduction budget, \ProvX significantly outperforms NoDoze and DepImpact in localization certainty, achieving a perfect Accuracy of 1.0000 and a Precision of 0.9798. Unlike traditional tools that provide a broader but less certain list of candidates—often leading to further analyst fatigue —\ProvX’s interventional reasoning provides a definitive and concise blocking strategy. This accuracy is paramount for automated response, where "cutting the wrong wire" can be as damaging as the attack itself.
\begin{table}[t]
		
	\centering
	\resizebox{0.35\textwidth}{!}{

			\begin{tabular}{@{}c|c|ccc@{}}
				\toprule
				\multirow{2}{*}{\begin{tabular}[c]{@{}c@{}}\textbf{GNN}\\ \textbf{Type}\end{tabular}} & \multirow{2}{*}{\begin{tabular}[c]{@{}c@{}}\textbf{DARPA}\\ \textbf{OpTC}\end{tabular}} & \multicolumn{3}{c}{\textbf{DARPA TC E3}} \\ \cmidrule(l){3-5} 
				&  & Cadets & Theia & Trace \\ \midrule
				GCN & 0.1856 & 0.3943 & 0.3197 & 0.3515 \\ \midrule
				GAT & 0.5926 & 0.4534 & 0.4356 & 0.4679 \\ \midrule
				GraphSAGE & 0.7695 & 0.4877 & 0.4556 & 0.5751 \\ \bottomrule
			\end{tabular}

	}
	\caption{Mitigation Efficacy of DARPA OpTC and TC-E3 on GNNs, with $K$=10 \& 30.}
\label{pn_all}
\end{table}

\begin{figure}[t]
	\centering
	\includegraphics[width=0.485\textwidth]{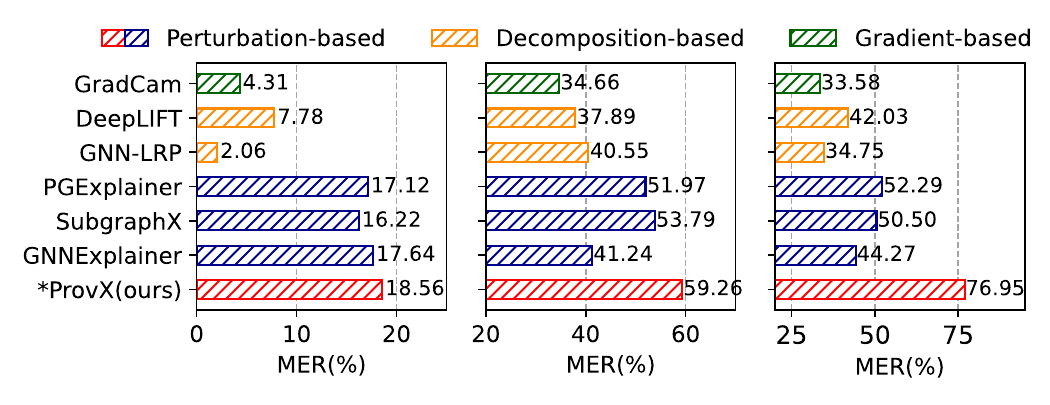}
	
	\caption{\ProvX’s MER(\%) performance on DARPA OpTC, from left to right are the results of \textbf{GCN}, \textbf{GAT}, and \textbf{GraphSAGE}. \ProvX's component is also a perturbation-based scheme, but is the only counterfactual-based explainer among these works.}
	\label{fig:pn_sota}
	
\end{figure}

\subsection{Blocking Effectiveness \textit{(RQ2)}}\label{RQ2}

RQ1 validates the localization ability of our framework, i.e., how well the edges localized by \ProvX correlate with real-world attack steps. However, from a security operations perspective, a more critical question remains: are these localized edges necessary for the detector's original alert? An edge that overlaps with attack ground truth is not necessarily the decisive evidence used by the GNN detector. Therefore, we further evaluate whether intervening on the top-$K$ core edge list is sufficient to flip the original detector prediction. Unlike methods that use Fidelity+ with surrogate models \cite{mukherjee2023interpreting}, we perform this validation directly on the fixed GNN detector. Inspired by concepts from causal inference \cite{tan2022learning, pearl2009causal, chu2024graph, tan2021counterfactual}, we introduce the \textbf{Mitigation Efficacy Rate (MER)} to measure the model-level blocking efficacy of the core edge list.

Let an alert subgraph be $G_k=(A_k, X_k)$, where $A_k$ is the adjacency matrix and $X_k$ is the node feature matrix. Given the top-$K$ core edge set $\tilde{G}^*_k$ output by \ProvX, we define a simulated intervention operator $\mathcal{I}_K(\cdot)$ that removes these selected edges from $A_k$ and suppresses the features of their incident endpoint nodes in $X_k$. The resulting intervened graph is then fed back into the same detector without retraining. MER is defined as follows:
\begin{equation}
	\text{MER} = \frac{1}{N} \sum_{k=1}^{N} p_k, \ \ \
	p_k =
	\begin{cases}
		1, & \text{if } f(\mathcal{I}_K(G_k)) \neq f(G_k), \\
		0, & \text{otherwise.}
	\end{cases}
\end{equation}
Here, $f(G_k)$ is the detector's original prediction on the alert subgraph, and $f(\mathcal{I}_K(G_k))$ is the prediction after applying the simulated structural-attribute intervention. A high MER indicates that the selected edge set and its local endpoint evidence are necessary for the detector to maintain its malicious judgment. MER is therefore a model-level metric: it evaluates whether the recommended core edges can flip the detector's alert under a controlled intervention, rather than proving that the corresponding host-level enforcement action is always safe or sufficient in deployment.

Table \ref{pn_all} presents the MER results of \ProvX on all datasets, while Figure \ref{fig:pn_sota} compares \ProvX with baseline algorithms under the same intervention protocol. We observe that, with $K=10$ ($K=30$ for DARPA TC E3), \ProvX achieves an MER greater than 18\% on all datasets, even achieves average MER of 51.59\% in OpTC. In other words, on the test set, the top-$K$ core edge list localized by \ProvX flips over half of the original detector alerts under the simulated intervention. Compared with fact-based baselines, \ProvX achieves higher MER across the three GNN backbones because it directly optimizes for a minimal counterfactual intervention that changes the detector's prediction, rather than only identifying subgraphs correlated with the original prediction.

\subsection{Cost-Benefit Analysis \textit{(RQ3)}}

The core motivation for our work is the alert fatigue crisis, characterized by voluminous, coarse-grained alerts that lack the prescriptive detail needed for response. While SOTA PIDS may correctly flag a subgraph as malicious, they often output tens of thousands of unprioritized system entities, imposing an immense review burden on analysts. In the context of defense, the critical trade-off is between Mitigation Precision (the safety of the intervention) and Operational Response Cost (the analyst's effort). 

This RQ evaluates \ProvX's ability to provide high-precision blocking strategies at a minimal cost. To conduct a fair comparison across systems with different output granularities, we define two metrics: Mitigation Precision (Y-axis): The proportion of real attack steps within the output list. High precision is paramount for operational safety, ensuring that blocking actions do not inadvertently disrupt benign business processes. Operational Response Cost (X-axis): The average number of entities (nodes or edges) an analyst must review to execute a blocking action, plotted on a logarithmic scale. For \ProvX, this cost is fixed at $K=30$ edges, representing a targeted intervention.

\begin{figure}[t]
	\centering
	\includegraphics[width=0.35\textwidth]{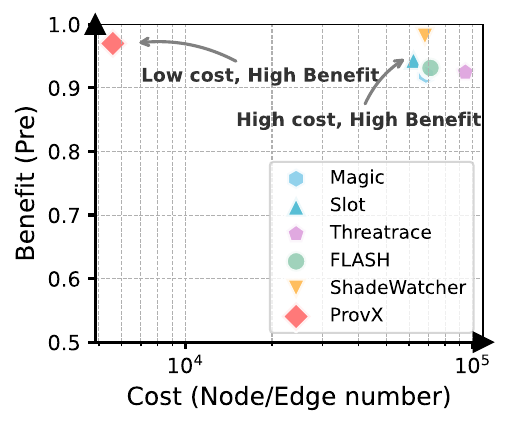}
	
	\caption{Cost-Benefit analysis of \ProvX against SOTA PIDS\cite{jia2024magic,qiao2024slot,rehman2024flash,zengy2022shadewatcher,wang2022threatrace} in DARPA E3 Trace. The X-axis represents the \textbf{Operational Cost} (average entities an analyst must review per alert). The Y-axis represents the \textbf{Benefit}. \ProvX is the only framework in the ideal \textit{Low Cost, High Benefit} quadrant.}
	\label{fig:cost_benefit}
	
\end{figure}

The results are shown in Figure \ref{fig:cost_benefit}. We observe a stark contrast between SOTA PIDS and \ProvX. Raw PIDS alerts reside in a High Cost, Low Precision category regarding specific entity-level actions; while they correctly identify the presence of a threat, their massive output (62,125 to 95,000 entities) is saturated with benign background noise, making the precision of any individual blocking action near zero without manual triage. Investigating such a volume would take a 10-analyst team roughly 6.9 to 11.2 months per alert, which is operationally impossible for active defense. In contrast, \ProvX resides in the ideal Low Cost, High Benefit quadrant. It achieves a near-perfect Mitigation Precision of 97.98\% while reducing the operational cost by over 89\% compared to investigating raw alerts. 

%This indicates that almost every "wire" \ProvX suggests cutting is indeed a critical attack step. This experiment demonstrates that \ProvX is the first framework to break the traditional SecOps trade-off, successfully transforming complex, high-cost alerts into a safe, high-precision, and immediately actionable blocking strategy.

\subsection{Hyperparameter Investigation \textit{(RQ4)}}
\label{RQ4}
\begin{figure}[t]
	\centering
	\includegraphics[width=0.48\textwidth]{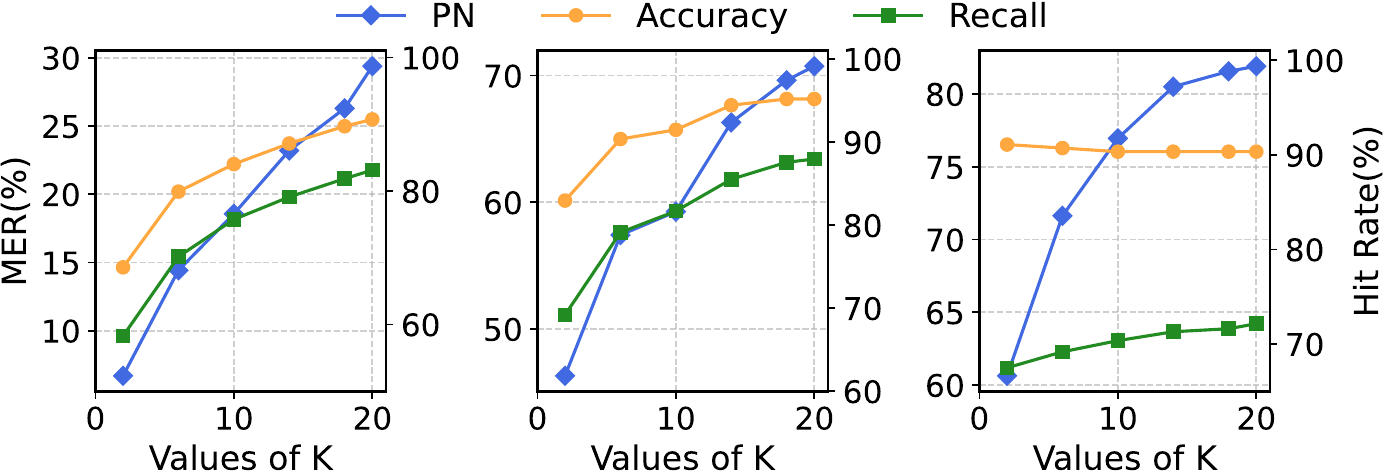}
	%\vspace{-0.1in}
	\caption{\ProvX's MER(\%) (left y-axis) and Hit Rate(\%) (right y-axis) as $K$ varies, from left to right are the results of GCN, GAT and GraphSAGE.}
	% \Description{A figure illustrating four scenes.}
	\label{fig:k_all}
	% \vspace{-0.1in}
\end{figure}

We investigate the impact of several different hyperparameters on the explanation performance of \ProvX:

\noindent \textbf{Response Budget \textbf{$K$}:} $K$ defines the number of edges included in the final core edge list, making it a direct proxy for the operational cost of analysis. We evaluate $K$ on the OpTC dataset, varying its value from 1 to 20 (Figure \ref{fig:k_all}). Increasing the $K$ budget directly improves the MER. On GraphSAGE, when $K=20$, the MER can reach 81.91\%. However, $K$ should not be excessively large, as this would diminish the significance of mitigation; removing a large portion of a graph will obviously cause a prediction flip.
% Furthermore, increasing $K$ also increases the Accuracy and Recall of the ground truth evaluation. 

\begin{figure}[t] 
	\centering 
	\subfigure[Mitigation Efficacy Rate]{ % 建议为子图添加标题
		\includegraphics[width=0.256\textwidth]{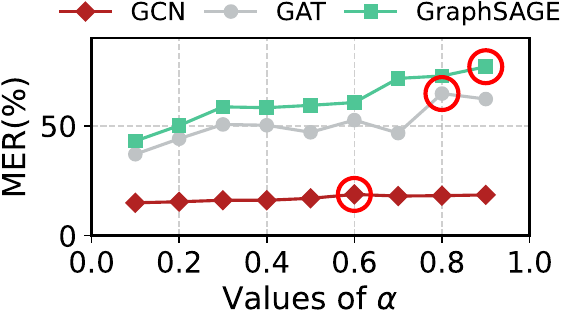}
		\label{fig:alpha_pn}
	}%
	% \hspace*{-5.5pt}%
	\subfigure[F1-score]{
		\includegraphics[width=0.17\textwidth]{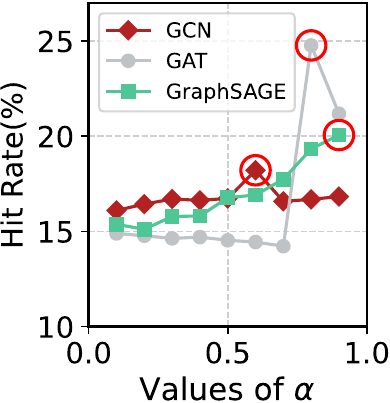}
		\label{fig:alpha_f1}
	}%
	\caption{MER and $F_1$-score of \ProvX when the $\alpha$ changes.}
	\label{fig:hy_alpha}
\end{figure}

\noindent \textbf{Counterfactual Trade-off Parameter $\alpha$:} The parameter $\alpha$ balances the predict flip loss ($\mathcal{L}_{\textbf{Pred}}$) and the mask distance Loss ($\mathcal{L}_{\textbf{dist}}$). The former aims to neutralize the alert, while the latter aims to minimize the blocking cost. The experimental results are shown in Figure \ref{fig:hy_alpha}. Figure \ref{fig:alpha_pn} displays the change in MER with respect to $\alpha$, and Figure \ref{fig:alpha_f1} shows the change in F1-score. On the three GNN detectors, the optimal $\alpha$ values are 0.6, 0.8, and 0.9, respectively, validated by both MER and F1-score. When $\alpha$ exceeds the optimal point, both metrics decline. 
%This demonstrates that the algorithm, by trying too hard to neutralize the alert, disrupts the balance and fails to find a solution at a minimal cost.

\begin{figure}[t] % 建议为figure环境加上位置参数，如htbp
	\centering 
	\subfigure[Solidification ratio study (while $\gamma_S=0.6$)]{ % 建议为子图添加标题
		\includegraphics[width=0.22\textwidth]{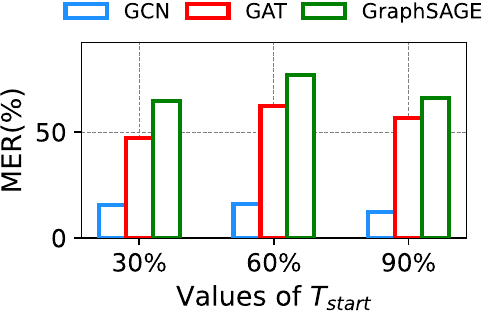}
		\label{fig:Ts}
	}
	% \hspace*{-5.5pt}%
	\subfigure[Solidification factor study (while $T_{start}=0.6$)]{
		\includegraphics[width=0.22\textwidth]{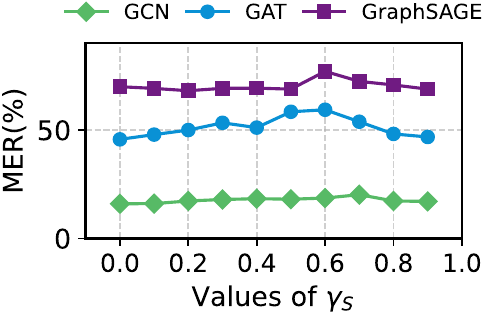}
		\label{fig:gammaS}
	}
	\caption{MER of \ProvX when the $T_{start}$ and $\gamma_S$ changes.}
	\label{fig:hy_Ts_sigmas}
\end{figure}

\noindent \textbf{Solidification Start Ratio $T_{start}$ \& Penalty Strength $\gamma_S$:} Solidifying the localization results is one of the core features of \ProvX. The $T_{start}$ ratio determines when to activate the mechanism, while the $\gamma_S$ strength determines the penalty for deviating from already-confident judgments. Figure \ref{fig:hy_Ts_sigmas} shows the analysis. Figure \ref{fig:Ts} shows the trend of MER as $T_{start}$ varies. A value of 0.6 provides the best balance, allowing sufficient exploratory time before the penalty phase begins. Figure \ref{fig:gammaS} shows the trend of MER as $\gamma_S$ varies. When $\gamma_S$ is 0 (no penalty), performance drops significantly. The best performance is achieved when $\gamma_S$ is 0.7, 0.6, and 0.6 for GCN, GAT, and GraphSAGE, respectively.

\subsection{System Overhead \textit{(RQ5)}}
\label{RQ5}

\begin{table}[t]

\centering
\resizebox{0.45\textwidth}{!}{

	\begin{tabular}{@{}cccccccc@{}}
	\toprule
	\multicolumn{2}{c}{\multirow{3}{*}{\textbf{Phase}}}             & \multicolumn{3}{c}{\textbf{Time consumption (s)}}                 & \multicolumn{3}{c}{\begin{tabular}[c]{@{}c@{}}\textbf{Peak Memory} \\ \textbf{consumption (MB)}\end{tabular}} \\ \cmidrule(l){3-8} 
	\multicolumn{2}{c}{}                                            & GCN & GAT  & \begin{tabular}[c]{@{}c@{}}Graph\\ SAGE\end{tabular} & GCN           & GAT          & \begin{tabular}[c]{@{}c@{}}Graph\\ SAGE\end{tabular}         \\ \midrule
	\multicolumn{2}{c}{Generation}                          & 113 & 174  & 140                                                  & 1377          & 1405         & 1379                                                         \\ \midrule
	\multicolumn{1}{c|}{\multirow{2}{*}{Mitigation}} & Training   & 733 & 1196 & 994                                                  & 1491          & 1519         & 1391                                                         \\ \cmidrule(l){2-8} 
	\multicolumn{1}{c|}{}                              & Evaluating & 3   & 4    & 4                                                    & 1541          & 1561         & 1461                                                         \\ \bottomrule
	\end{tabular}
}
\caption{Overhead analysis of the \ProvX framework, detailing the \textbf{Phase I (Detection)} and \textbf{Phase II (Mitigation)} components.}

\label{overhead}
\end{table}

%\ProvX is a two-stage PIDS framework, and its complete workflow includes the training for Phase I (Alert Generation) and the training and evaluation for Phase II (Alert Refinement). We test the time and memory overhead for three types of GNN cores. Notably, considering that most real-world threat analysis scenarios lack GPU support, we conduct our evaluation in a CPU environment.
%
%The experimental results are shown in TABLE \ref{overhead}. As can be seen, the training for Phase I.Alert Generation) (Detection Training) does not incur excessive overhead. For the Phase II.Alert Refinement) training (Refinement Training), even the longest case (using GAT) requires less than 20 minutes, and the average peak memory usage is only 1.43 GB, which can be easily handled by a standard host machine. Furthermore, the final refinement evaluation (Refinement Evaluating) can be completed in just 3-4 seconds.
%
%Overall, \ProvX demonstrates excellent performance in terms of time and memory overhead. Its low hardware requirements significantly reduce the barrier and cost for enterprises to deploy advanced alert refinement capabilities, allowing for easy integration into existing standard server architectures. Security teams can leverage it to rapidly iterate on their alert analysis models based on newly emerging attack patterns.

\ProvX is an defense framework whose workflow encompasses two phase. We evaluate its performance across three GNN backbones in a CPU-only environment, reflecting realistic SecOps scenarios where dedicated GPU support for forensic analysis is often unavailable. As shown in Table \ref{overhead}, the detection training in Phase I is highly efficient. For Phase II, the average peak memory remains remarkably low at 1.43 GB. Critically, once the strategy is synthesized, the final mitigation evaluation (identifying the specific choke points) is completed in just 3-4 seconds.

Overall, \ProvX demonstrates high operational feasibility with minimal hardware requirements. We argue that the minute-level synthesis time is a justifiable trade-off for high-precision, analyst-facing blocking candidates, especially when compared to the hours or days an analyst would otherwise spend manually investigating voluminous raw alerts. By reducing the computational barrier, \ProvX allows security teams to rapidly deploy and iterate on defense strategies within existing standard server architectures, providing a practical pathway for timely alert analysis and blocking-oriented response.
\begin{figure}[t] % 建议为figure环境加上位置参数，如htbp
	\centering 
	\subfigure[Localization performance before and after adding adversarial structure.]{ % 建议为子图添加标题
		\includegraphics[width=0.217\textwidth]{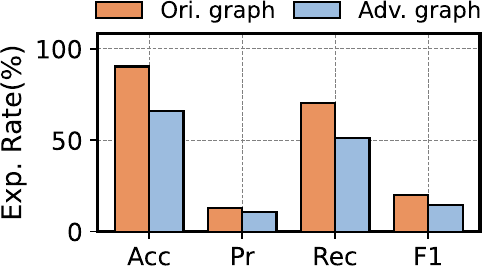}
		\label{fig:ad_exp}
	}
	% \hspace*{-3.5pt}%
	\subfigure[Detection performance before and after feedback.]{
		\includegraphics[width=0.22\textwidth]{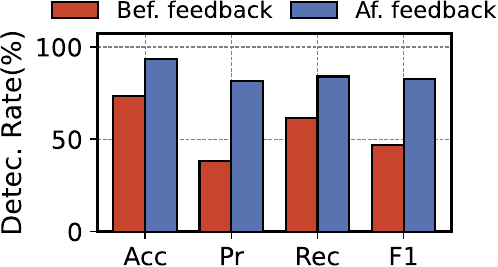}
		\label{fig:ad_ids}
	}
%	\caption{\ProvX finds the key structures that evade model detection and feeds back to the security model to enhance adversarial robustness.}
	\caption{Demonstration of the \textbf{Feedback Loop}. (a) \ProvX's \textbf{localization performance} is impacted by adversarial structures. (b) However, feeding these \textbf{localized analysis results} back for retraining significantly enhances the detector's adversarial robustness.}
	\label{fig:ad}
\end{figure}

%\begin{figure*}[t]
%	\centering
%	\includegraphics[width=0.99\textwidth]{figure/case_study.pdf}
%	%\vspace{-0.1in}
%	\caption{An attack case from the DARPA E3 Theia dataset. We construct a provenance graph from the audit log, which includes processes, files, and sockets. The figure shows four possible ways to perturb \ProvX.}
%	% \Description{A figure illustrating four scenes.}
%	\label{fig:case_study}
%	% \vspace{-0.1in}
%\end{figure*}

\subsection{Adversarial Analysis \textit{(RQ6)}}
\label{RQ6}

In this section, we provide a preliminary evaluation of the feedback idea described in Section \ref{feedback}. Starting from existing malicious provenance subgraphs, we inject benign-looking structures following the mimicry attack setting of \cite{goyal2023sometimes} and retain the generated samples that evade the detector as $G_{adv}$. We then apply \ProvX with the reversed objective: instead of finding structures whose removal flips a malicious prediction to benign, it finds structures whose removal flips the false-negative prediction from benign back to malicious.

The localized structure is treated as candidate evasion-associated context. After this step, we construct $G_{left}$ by removing the localized structure from $G_{adv}$. We feed both $G_{adv}$ and $G_{left}$, with malicious labels, back into detector training. We then report two effects: (1) how adversarial camouflage affects \ProvX's localization performance, and (2) how feedback changes the detector's performance on adversarial samples, as shown in Figure \ref{fig:ad_exp} and Figure \ref{fig:ad_ids}.

As shown in Figure \ref{fig:ad_exp}, adding adversarial structures decreases localization performance because the injected benign-looking context changes the detector's decision boundary and weakens the visibility of the attack core. Nevertheless, \ProvX still identifies candidate structures associated with the evasion. After feedback training, the detector's performance on adversarial samples improves: Accuracy, Precision, Recall, and F1 improve by 20.1\%, 43.7\%, 22.2\%, and 35.5\%, respectively (Figure \ref{fig:ad_ids}). These results suggest that \ProvX's localized feedback can provide useful training signals for detector hardening, while a complete feedback framework remains future work.
\section{Discussion \& Future Work}
\label{discussion}

%\noindent \textbf{Scalability and Real-Time Performance.} The search for a minimal counterfactual explanation, as employed by \ProvX, is fundamentally a computationally intensive process, often approximating an NP-hard problem. Our current implementation operates by searching for the minimal structural perturbation via a computationally-intensive continuous optimization process. While effective for offline forensic analysis, this presents a significant computational bottleneck. In a real-time PIDS context, such as a SecOps Center managing a high-throughput event stream, the latency required to generate an explanation for every alert could be prohibitive. This limits the immediate applicability of \ProvX in time-sensitive, high-load environments.
\noindent \textbf{Positioning and Industrial Feasibility.}
\ProvX is designed as a forensic decision-support tool rather than a real-time, in-line enforcement system. Its optimization is intended to operate post-detection, transforming voluminous, unactionable alerts into analyst-reviewable candidate choke points. In practical SecOps, the bottleneck is often not only sub-second detection latency, but the long manual effort required to investigate complex attack chains. By providing compact blocking-oriented guidance in minutes, \ProvX helps bridge the gap between threat discovery and response planning. However, mapping a provenance edge to a safe host-level action, such as process termination, file quarantine, syscall blocking, or network restriction, remains deployment-specific and must account for collateral cost and business context.

 \noindent \textbf{Alert-Subgraph Acquisition.}
 \label{subgraph}
\ProvX assumes that an alert-associated provenance subgraph is available as input. In our experiments, we approximate this SOC input by using a Louvain-based procedure to obtain evaluation subgraphs from DARPA provenance graphs, followed by a GNN-based detector to identify malicious alerts. This setting enables controlled evaluation on public datasets, but it does not fully reproduce how alerts are generated, correlated, and expanded in a live SOC. In deployment, the input subgraph may come from mature PIDS, EDR/SIEM alarms, network indicators mapped to host events, or analyst-guided provenance expansion. 
We also observe that the quality of this upstream subgraph acquisition can noticeably affect downstream blocking-oriented refinement. If a subgraph is too small, the relevant attack dependency may be split across windows or communities; if it is too large, benign background behavior may dilute the structural signal and make core-edge localization harder. Thus, obtaining alert subgraphs with stronger structural discriminability is an important direction for improving \ProvX. Beyond the Louvain-based instantiation used in this paper, preliminary exploration with time-window-based subgraphs and ego-network expansion around suspicious seed entities also produced meaningful inputs for the blocking module. We therefore view Phase I as a replaceable alert-acquisition layer: the current counterfactual blocking module can benefit from more mature SOC subgraph construction strategies, and designing such strategies is part of our future work.

\noindent \textbf{Applicability to Non-Differentiable PIDS.} The current design of \ProvX is tailored for explaining GNNs, which are (typically) differentiable models. Our methodology, which may rely on probing model-internal states or structural sensitivities, does not directly apply to the wide array of non-differentiable PIDS currently in use, such as those based on decision forests, isolation forests, or complex rule-based expert systems. This limits the generalizability of \ProvX as a universal explainer for all provenance-based detection. To address the aforementioned limitation, a key future direction is to adapt \ProvX into a truly model-agnostic framework. 
We plan to explore techniques inspired by LIME or SHAP, which treat the detector as a black box and query it to build a local, interpretable approximation. This would allow \ProvX to be integrated as a modular explanation component for a much wider range of existing PIDS.

\noindent \textbf{Formalizing the Alert-Feedback Loop.} In the main text, we present the feedback loop as a preliminary extension of \ProvX. A complete deployment-ready feedback framework still requires further formalization and implementation. Future work should address several key questions: how analyst feedback should be encoded, how feedback samples should be weighted during retraining or fine-tuning, and how the detector can benefit from new adversarial examples without catastrophic forgetting. Exploring active learning or reinforcement learning from human feedback may be a promising direction for systematically improving both detector robustness and explanation quality.

\section{Conclusion}
\label{sec:conclusion}

We propose \ProvX, an active APT defense framework designed to bridge the critical operational gap between threat detection and mitigation. \ProvX employs an interventional reasoning technique based on counterfactual logic. This method reformulates the search for critical attack edges into a continuous, differentiable optimization task. By maximizing a prediction flip loss while minimizing a mask distance Loss, and incorporating a novel staged solidification strategy, \ProvX effectively isolates the minimal structural subset—the "fuses" that are most necessary for the detector's malicious judgment and should be prioritized for blocking-oriented response.

Evaluations on real-world APT datasets demonstrate that \ProvX achieves good performance, ensuring that blocking recommendations are both highly relevant to actual attack steps and safe for system continuity. This work also validates a preliminary closed-loop feedback framework, positioning \ProvX as a forensic decision-support tool that helps make existing PIDS outputs more actionable while reducing analyst burden.

%\cleardoublepage
%\appendix

\section*{Acknowledgment}
%Development Program of China (2023YFB2904000),  National Natural Science Foundation of China under Grant (No. 62272370), Natural Science Basic Research Program of Shaanxi (No. 2025JC-JCQN-073), china Higher Education Institutions Industry-University-Research Innovation Fund (Project No. 2024IT095), Qinchuangyuan Scientist + Engineer Team Program of Shaanxi (No. 2024QCY-KXJ-149), Songshan Laboratory (No. 241110210200),Fundamental Research Funds for the Central Universities(No. YJSJ26003), National Research Foundation, Singapore, and DSO National Laboratories under the AI Singapore Programme (AISG Award No: AISG2-GC-2023-008), the National Research Foundation, Singapore, and the Cyber Security Agency under its National Cybersecurity R\&D Programme (NCRP25-P04-TAICeN), the National Research Foundation, Prime Minister’s Office, Singapore under Campus for Research Excellence and Technological Enterprise (CREATE) programme, and Ripple under University Blockchain Research Initiative (UBRI)\cite{feng2022university}.

We thank anonymous reviewers for their valuable comments.
This work was supported in part by the National Key Research and Development Program of China (2023YFB2904000), the National Natural Science Foundation of China under Grant (No. 62272370), the Natural Science Basic Research Program of Shaanxi (No. 2025JC-JCQN-073), China Higher Education Institutions Industry-University-Research Innovation Fund (Project No. 2024IT095), Qinchuangyuan Scientist + Engineer Team Program of Shaanxi (No. 2024QCY-KXJ-149), Songshan Laboratory (No. 241110210200), Fundamental Research Funds for the Central Universities (No. YJSJ26003), National Research Foundation, Singapore, and DSO National Laboratories under the AI Singapore Programme (AISG Award No. AISG2-GC-2023-008), the National Research Foundation, Singapore, and the Cyber Security Agency under its National Cybersecurity R\&D Programme (NCRP25-P04-TAICeN), the National Research Foundation, Prime Minister's Office, Singapore under Campus for Research Excellence and Technological Enterprise (CREATE) programme, Ripple under University Blockchain Research Initiative (UBRI)\cite{feng2022university}, the National Natural Science Foundation of China under Grant 62402202, and project ZR2024QF036 supported by Shandong Provincial Natural Science Foundation.

\section*{Ethical Considerations} 
%\textbf{We attest that the research team has carefully reviewed the conference ethics guidelines and conducted this study in accordance with established ethical standards. Our research utilizes publicly available datasets. These datasets consist of anonymized system audit logs collected in controlled environments. No personally identifiable information (PII) or sensitive user data was collected or processed during this study. The proposed framework, \ProvX, is designed for defensive purposes, specifically to improve the actionability of intrusion detection systems through attack blocking. Our experiments were conducted on isolated research datasets and did not involve any intervention in live production systems or the exploitation of unpatched vulnerabilities.  While we acknowledge that counterfactual reasoning can theoretically be used to explore evasion paths, our work focuses strictly on defense-centric "choke point" localization to empower security analysts. We believe the defensive benefits—neutralizing APT threats and reducing analyst fatigue—significantly outweigh the potential for misuse. The research team's future plans remain committed to advancing ethical, proactive defense technologies.}

This section examines the ethical implications of \ProvX by identifying affected stakeholders, assessing potential benefits and harms, outlining safeguards for responsible deployment, and explaining our decision to conduct and publish this research.

\noindent\textbf{Stakeholders.}
The stakeholders affected by this research include system administrators and SOC analysts who may deploy \ProvX, end users whose activities may be captured in provenance logs, organizations and system owners protected by PIDS, security tool vendors who may adopt similar techniques, the broader security research community, adversaries whose attack strategies may be disrupted or informed by this work, and society at large, including users of critical infrastructure, financial systems, and government services targeted by APT campaigns.

\noindent\textbf{Positive Impacts.}
\ProvX is designed to improve APT defense by moving beyond detection-only alerts toward \textit{analyst-reviewable blocking guidance}. For SOC analysts, \ProvX can reduce the cognitive burden of incident response by identifying a compact set of provenance edges that are important to the detector's malicious judgment. For organizations, this can support faster containment and reduce adversary dwell time. For end users and the public, stronger APT defenses can help protect the confidentiality, integrity, and availability of systems that process sensitive data and support critical services. For the research community, this work contributes a counterfactual reasoning framework and the Mitigation Efficacy Rate (MER) metric for evaluating \textit{model-level blocking efficacy}.

\noindent\textbf{Potential Negative Impacts.}
Despite its defensive intent, \ProvX may introduce several risks if deployed carelessly. First, acting on blocking recommendations in production may disrupt benign activity. A false positive mitigation could terminate critical processes, sever legitimate network connections, or block access to essential files, causing service outages, lost work, or cascading operational failures. Second, because \ProvX prioritizes a subset of alert edges, analysts may over-focus on the recommended core edge list and overlook genuine threats outside the prioritized set. This could introduce new false negatives at the triage stage, even when the underlying PIDS has detected the broader suspicious activity. Third, \ProvX's counterfactual reasoning pipeline introduces processing overhead, which may limit usefulness in time-critical response settings or large-scale deployments. Fourth, provenance logging systems such as Linux Audit, ETW, and CamFlow collect fine-grained process, file, and network activity. Such logs may raise privacy concerns if repurposed for employee surveillance or other non-security uses. Finally, the counterfactual reasoning methodology has \textit{dual-use potential}: an adversary with access to comparable provenance data and detector behavior might use similar reasoning to make attacks more resilient to edge-level mitigations.

\noindent\textbf{Mitigations.}
We mitigate these risks in several ways. Our evaluation uses only publicly available DARPA TC and OpTC benchmark datasets collected in controlled environments; we do not collect, store, or analyze real user data or personally identifiable information. For real-world deployment, organizations should apply \textit{data minimization}, log rotation, aggregation or anonymization where feasible, clear retention limits, and access controls for provenance logs. Operationally, \ProvX should be deployed as a \textit{decision-support tool} rather than a fully automated response system. Its recommendations should be reviewed by human analysts before any production mitigation is executed, and automated enforcement should first be evaluated in dry-run or shadow mode. To reduce false-negative risk from triage, organizations should preserve the full PIDS alert corpus alongside \ProvX's prioritized subset and periodically review deprioritized alerts. Future work will also investigate streaming and incremental computation to reduce latency. We acknowledge the dual-use nature of this research, but note that \ProvX primarily benefits defenders because it assumes access to detailed provenance logs and a trained detection model, which are typically available to the defending organization.

\noindent\textbf{Unmitigated Risks.}
Some risks cannot be fully eliminated. Provenance logging infrastructure may still be repurposed for surveillance despite our recommendations. \ProvX's prioritization may still cause analysts to miss threats outside the recommended core edge list. Counterfactual reasoning may remain too slow for some time-critical scenarios. Publication of the method may also inform adversarial evasion strategies. We consider these risks inherent to security research that produces \textit{dual-use knowledge}.

\noindent\textbf{Decision to Conduct and Publish.}
Our decision to conduct and publish this research is guided by \textit{beneficence}, \textit{respect for persons}, \textit{justice}, and \textit{respect for law and public interest}. APT attacks cause substantial harm to governments, enterprises, and individuals, while current incident response remains heavily manual and time-consuming. \ProvX aims to improve this situation by providing analysts with actionable, human-reviewed guidance. The study does not involve human subjects and relies only on public benchmark datasets, so no informed consent issue arises in our experiments. For deployment, however, organizations should notify users about provenance logging and establish transparent data governance policies. We also recognize that organizations need provenance logging infrastructure and GNN-based PIDS to benefit from \ProvX, which may limit accessibility for some resource-constrained organizations. We comply with applicable dataset terms of use. Overall, we believe the defensive value of improving APT response outweighs the residual dual-use and operational risks, particularly because open publication enables safer follow-on work.

%\cleardoublepage

\section*{Open Science}
\textbf{To support future research, we provide an accompanying artifact repository\cite{provx-artifact} for ProvX. The artifact contains the code implementation, including the GNN detector, the ProvX interventional mask learning module, the staged solidification procedure, and evaluation utilities. }
	
%	.
%	
%\textbf{The repository contains the following artifacts: The complete source code of our two-stage framework, including the interventional mask learning and staged solidification modules. Scripts and configurations required to reproduce the localization accuracy, mitigation efficacy and hypermaters study. }
% optional clearing of the page
%\cleardoublepage

\bibliographystyle{plainurl}
\bibliography{reference}

\begin{thebibliography}{10}

\bibitem{ATTCK}
Adversarial tactics, techniques and common knowledge.
\newblock \url{https://attack. mitre.org/wiki/Main Page.}

\bibitem{DARPA_OpTC}
Darpa operationally transparent cyber (optc) data release.
\newblock \url{https://github.com/FiveDirections/OpTC-data}.
\newblock 2019.

\bibitem{DARPA}
Darpa transparent computing program engagement 3 data release.
\newblock \url{https://github.com/darpa-i2o/ Transparent-Computing}.
\newblock 2020.

\bibitem{provx-artifact}
{ProvX Artifact}.
\newblock URL: \url{https://zenodo.org/records/20310415}, \href
  {https://doi.org/10.5281/zenodo.20310415}
  {\path{doi:10.5281/zenodo.20310415}}.

\bibitem{alsaheel2021atlas}
Abdulellah Alsaheel, Yuhong Nan, Shiqing Ma, Le~Yu, Gregory Walkup, Z~Berkay
  Celik, Xiangyu Zhang, and Dongyan Xu.
\newblock $\{$ATLAS$\}$: A sequence-based learning approach for attack
  investigation.
\newblock In {\em 30th USENIX security symposium (USENIX security 21)}, pages
  3005--3022, 2021.

\bibitem{bates2015trustworthy}
Adam Bates, Dave~Jing Tian, Kevin~RB Butler, and Thomas Moyer.
\newblock Trustworthy $\{$Whole-System$\}$ provenance for the linux kernel.
\newblock In {\em 24th USENIX Security Symposium (USENIX Security 15)}, pages
  319--334, 2015.

\bibitem{blondel2008fast}
Vincent~D Blondel, Jean-Loup Guillaume, Renaud Lambiotte, and Etienne Lefebvre.
\newblock Fast unfolding of communities in large networks.
\newblock {\em Journal of statistical mechanics: theory and experiment},
  2008(10):P10008, 2008.

\bibitem{cheng2024kairos}
Zijun Cheng, Qiujian Lv, Jinyuan Liang, Yan Wang, Degang Sun, Thomas Pasquier,
  and Xueyuan Han.
\newblock Kairos: Practical intrusion detection and investigation using
  whole-system provenance.
\newblock In {\em 2024 IEEE Symposium on Security and Privacy (SP)}, pages
  3533--3551. IEEE, 2024.

\bibitem{chu2024graph}
Zhaoyang Chu, Yao Wan, Qian Li, Yang Wu, Hongyu Zhang, Yulei Sui, Guandong Xu,
  and Hai Jin.
\newblock Graph neural networks for vulnerability detection: A counterfactual
  explanation.
\newblock In {\em Proceedings of the 33rd ACM SIGSOFT International Symposium
  on Software Testing and Analysis}, pages 389--401, 2024.

\bibitem{fang2022back}
Pengcheng Fang, Peng Gao, Changlin Liu, Erman Ayday, Kangkook Jee, Ting Wang,
  Yanfang~Fanny Ye, Zhuotao Liu, and Xusheng Xiao.
\newblock $\{$Back-Propagating$\}$ system dependency impact for attack
  investigation.
\newblock In {\em 31st USENIX security symposium (USENIX Security 22)}, pages
  2461--2478, 2022.

\bibitem{feng2022university}
Yebo Feng, Jiahua Xu, and Lauren Weymouth.
\newblock University blockchain research initiative (ubri): Boosting blockchain
  education and research.
\newblock {\em IEEE Potentials}, 41(6):19--25, 2022.

\bibitem{goyal2023sometimes}
Akul Goyal, Xueyuan Han, Gang Wang, and Adam Bates.
\newblock Sometimes, you aren't what you do: Mimicry attacks against provenance
  graph host intrusion detection systems.
\newblock In {\em 30th Network and Distributed System Security Symposium},
  2023.

\bibitem{goyal2024r}
Akul Goyal, Gang Wang, and Adam Bates.
\newblock R-caid: Embedding root cause analysis within provenance-based
  intrusion detection.
\newblock In {\em 2024 IEEE Symposium on Security and Privacy (SP)}, pages
  3515--3532. IEEE, 2024.

\bibitem{hanunicorn}
Xueyuan Han, Thomas Pasquier, Adam Bates, James Mickens, and Margo Seltzer.
\newblock Unicorn: Runtime provenance-based detector for advanced persistent
  threats.
\newblock In {\em In Network and Distributed System Security Symposium
  (NDSS’20).}, 2020.

\bibitem{han2021sigl}
Xueyuan Han, Xiao Yu, Thomas Pasquier, Ding Li, Junghwan Rhee, James Mickens,
  Margo Seltzer, and Haifeng Chen.
\newblock $\{$SIGL$\}$: Securing software installations through deep graph
  learning.
\newblock In {\em 30th USENIX Security Symposium (USENIX Security 21)}, pages
  2345--2362, 2021.

\bibitem{hassan2020tactical}
Wajih~Ul Hassan, Adam Bates, and Daniel Marino.
\newblock Tactical provenance analysis for endpoint detection and response
  systems.
\newblock In {\em 2020 IEEE Symposium on Security and Privacy (SP)}, pages
  1172--1189. IEEE, 2020.

\bibitem{hassan2019nodoze}
Wajih~Ul Hassan, Shengjian Guo, Ding Li, Zhengzhang Chen, Kangkook Jee, Zhichun
  Li, and Adam Bates.
\newblock Nodoze: Combatting threat alert fatigue with automated provenance
  triage.
\newblock In {\em network and distributed systems security symposium}, 2019.

\bibitem{hassan2020we}
Wajih~Ul Hassan, Ding Li, Kangkook Jee, Xiao Yu, Kexuan Zou, Dawei Wang,
  Zhengzhang Chen, Zhichun Li, Junghwan Rhee, Jiaping Gui, et~al.
\newblock This is why we can’t cache nice things: Lightning-fast threat
  hunting using suspicion-based hierarchical storage.
\newblock In {\em Proceedings of the 36th Annual Computer Security Applications
  Conference}, pages 165--178, 2020.

\bibitem{hossain2017sleuth}
Md~Nahid Hossain, Sadegh~M Milajerdi, Junao Wang, Birhanu Eshete, Rigel
  Gjomemo, R~Sekar, Scott Stoller, and VN~Venkatakrishnan.
\newblock Sleuth: Real-time attack scenario reconstruction from cots audit
  data.
\newblock In {\em 26th USENIX Security Symposium (USENIX Security 17)}, pages
  487--504, 2017.

\bibitem{hu2023interpreters}
Yutao Hu, Suyuan Wang, Wenke Li, Junru Peng, Yueming Wu, Deqing Zou, and Hai
  Jin.
\newblock Interpreters for gnn-based vulnerability detection: Are we there yet?
\newblock In {\em Proceedings of the 32nd ACM SIGSOFT International Symposium
  on Software Testing and Analysis}, pages 1407--1419, 2023.

\bibitem{inam2023sok}
Muhammad~Adil Inam, Yinfang Chen, Akul Goyal, Jason Liu, Jaron Mink, Noor
  Michael, Sneha Gaur, Adam Bates, and Wajih~Ul Hassan.
\newblock Sok: History is a vast early warning system: Auditing the provenance
  of system intrusions.
\newblock In {\em 2023 IEEE Symposium on Security and Privacy (SP)}, pages
  2620--2638. IEEE, 2023.

\bibitem{jia2024magic}
Zian Jia, Yun Xiong, Yuhong Nan, Yao Zhang, Jinjing Zhao, and Mi~Wen.
\newblock $\{$MAGIC$\}$: Detecting advanced persistent threats via masked graph
  representation learning.
\newblock In {\em 33rd USENIX Security Symposium (USENIX Security 24)}, pages
  5197--5214, 2024.

\bibitem{jiang2025orthrus}
Baoxiang Jiang, Tristan Bilot, Nour El~Madhoun, Khaldoun Al~Agha, Anis Zouaoui,
  Shahrear Iqbal, Xueyuan Han, and Thomas Pasquier.
\newblock Orthrus: Achieving high quality of attribution in provenance-based
  intrusion detection systems.
\newblock In {\em Security Symposium (USENIX Sec’25). USENIX}, 2025.

\bibitem{li2021hierarchical}
Zitong Li, Xiang Cheng, Lixiao Sun, Ji~Zhang, and Bing Chen.
\newblock A hierarchical approach for advanced persistent threat detection with
  attention-based graph neural networks.
\newblock {\em Security and Communication Networks}, 2021:1--14, 2021.

\bibitem{liu2018towards}
Yushan Liu, Mu~Zhang, Ding Li, Kangkook Jee, Zhichun Li, Zhenyu Wu, Junghwan
  Rhee, and Prateek Mittal.
\newblock Towards a timely causality analysis for enterprise security.
\newblock In {\em NDSS}, 2018.

\bibitem{luo2020parameterized}
Dongsheng Luo, Wei Cheng, Dongkuan Xu, Wenchao Yu, Bo~Zong, Haifeng Chen, and
  Xiang Zhang.
\newblock Parameterized explainer for graph neural network.
\newblock {\em Advances in neural information processing systems},
  33:19620--19631, 2020.

\bibitem{milajerdi2019poirot}
Sadegh~M Milajerdi, Birhanu Eshete, Rigel Gjomemo, and VN~Venkatakrishnan.
\newblock Poirot: Aligning attack behavior with kernel audit records for cyber
  threat hunting.
\newblock In {\em Proceedings of the 2019 ACM SIGSAC conference on computer and
  communications security}, pages 1795--1812, 2019.

\bibitem{milajerdi2019holmes}
Sadegh~M Milajerdi, Rigel Gjomemo, Birhanu Eshete, Ramachandran Sekar, and
  VN~Venkatakrishnan.
\newblock Holmes: real-time apt detection through correlation of suspicious
  information flows.
\newblock In {\em 2019 IEEE Symposium on Security and Privacy (SP)}, pages
  1137--1152. IEEE, 2019.

\bibitem{mukherjee2023interpreting}
Kunal Mukherjee, Joshua Wiedemeier, Tianhao Wang, Muhyun Kim, Feng Chen, Murat
  Kantarcioglu, and Kangkook Jee.
\newblock Interpreting gnn-based ids detections using provenance graph
  structural features.
\newblock {\em arXiv e-prints}, pages arXiv--2306, 2023.

\bibitem{pasquier2017practical}
Thomas Pasquier, Xueyuan Han, Mark Goldstein, Thomas Moyer, David Eyers, Margo
  Seltzer, and Jean Bacon.
\newblock Practical whole-system provenance capture.
\newblock In {\em Proceedings of the 2017 Symposium on Cloud Computing}, pages
  405--418, 2017.

\bibitem{pearl2009causal}
Judea Pearl.
\newblock Causal inference in statistics: An overview.
\newblock 2009.

\bibitem{pei2016hercule}
Kexin Pei, Zhongshu Gu, Brendan Saltaformaggio, Shiqing Ma, Fei Wang, Zhiwei
  Zhang, Luo Si, Xiangyu Zhang, and Dongyan Xu.
\newblock Hercule: Attack story reconstruction via community discovery on
  correlated log graph.
\newblock In {\em Proceedings of the 32Nd Annual Conference on Computer
  Security Applications}, pages 583--595, 2016.

\bibitem{qiao2024slot}
Wei Qiao, Yebo Feng, Teng Li, Zhuo Ma, Yulong Shen, JianFeng Ma, and Yang Liu.
\newblock Slot: Provenance-driven apt detection through graph reinforcement
  learning.
\newblock {\em arXiv preprint arXiv:2410.17910}, 2024.

\bibitem{rehman2024flash}
Mati~Ur Rehman, Hadi Ahmadi, and Wajih~Ul Hassan.
\newblock Flash: A comprehensive approach to intrusion detection via provenance
  graph representation learning.
\newblock In {\em 2024 IEEE Symposium on Security and Privacy (SP)}, pages
  139--139. IEEE Computer Society, 2024.

\bibitem{schnake2021higher}
Thomas Schnake, Oliver Eberle, Jonas Lederer, Shinichi Nakajima, Kristof~T
  Sch{\"u}tt, Klaus-Robert M{\"u}ller, and Gr{\'e}goire Montavon.
\newblock Higher-order explanations of graph neural networks via relevant
  walks.
\newblock {\em IEEE transactions on pattern analysis and machine intelligence},
  44(11):7581--7596, 2021.

\bibitem{selvaraju2017grad}
Ramprasaath~R Selvaraju, Michael Cogswell, Abhishek Das, Ramakrishna Vedantam,
  Devi Parikh, and Dhruv Batra.
\newblock Grad-cam: Visual explanations from deep networks via gradient-based
  localization.
\newblock In {\em Proceedings of the IEEE international conference on computer
  vision}, pages 618--626, 2017.

\bibitem{shen2018tiresias}
Yun Shen, Enrico Mariconti, Pierre~Antoine Vervier, and Gianluca Stringhini.
\newblock Tiresias: Predicting security events through deep learning.
\newblock In {\em Proceedings of the 2018 ACM SIGSAC Conference on Computer and
  Communications Security}, pages 592--605, 2018.

\bibitem{shrikumar2017learning}
Avanti Shrikumar, Peyton Greenside, and Anshul Kundaje.
\newblock Learning important features through propagating activation
  differences.
\newblock In {\em International conference on machine learning}, pages
  3145--3153. PMlR, 2017.

\bibitem{tan2022learning}
Juntao Tan, Shijie Geng, Zuohui Fu, Yingqiang Ge, Shuyuan Xu, Yunqi Li, and
  Yongfeng Zhang.
\newblock Learning and evaluating graph neural network explanations based on
  counterfactual and factual reasoning.
\newblock In {\em Proceedings of the ACM web conference 2022}, pages
  1018--1027, 2022.

\bibitem{tan2021counterfactual}
Juntao Tan, Shuyuan Xu, Yingqiang Ge, Yunqi Li, Xu~Chen, and Yongfeng Zhang.
\newblock Counterfactual explainable recommendation.
\newblock In {\em Proceedings of the 30th ACM International Conference on
  Information \& Knowledge Management}, pages 1784--1793, 2021.

\bibitem{tariq2025alert}
Shahroz Tariq, Mohan Baruwal~Chhetri, Surya Nepal, and Cecile Paris.
\newblock Alert fatigue in security operations centres: Research challenges and
  opportunities.
\newblock {\em ACM Computing Surveys}, 57(9):1--38, 2025.

\bibitem{wang2020you}
Qi~Wang, Wajih~Ul Hassan, Ding Li, Kangkook Jee, Xiao Yu, Kexuan Zou, Junghwan
  Rhee, Zhengzhang Chen, Wei Cheng, Carl~A Gunter, et~al.
\newblock You are what you do: Hunting stealthy malware via data provenance
  analysis.
\newblock In {\em NDSS}, 2020.

\bibitem{wang2022threatrace}
Su~Wang, Zhiliang Wang, Tao Zhou, Hongbin Sun, Xia Yin, Dongqi Han, Han Zhang,
  Xingang Shi, and Jiahai Yang.
\newblock Threatrace: Detecting and tracing host-based threats in node level
  through provenance graph learning.
\newblock {\em IEEE Transactions on Information Forensics and Security},
  17:3972--3987, 2022.

\bibitem{ying2019gnnexplainer}
Zhitao Ying, Dylan Bourgeois, Jiaxuan You, Marinka Zitnik, and Jure Leskovec.
\newblock Gnnexplainer: Generating explanations for graph neural networks.
\newblock {\em Advances in neural information processing systems}, 32, 2019.

\bibitem{yuan2021explainability}
Hao Yuan, Haiyang Yu, Jie Wang, Kang Li, and Shuiwang Ji.
\newblock On explainability of graph neural networks via subgraph explorations.
\newblock In {\em International conference on machine learning}, pages
  12241--12252. PMLR, 2021.

\bibitem{zengy2022shadewatcher}
Jun Zengy, Xiang Wang, Jiahao Liu, Yinfang Chen, Zhenkai Liang, Tat-Seng Chua,
  and Zheng~Leong Chua.
\newblock Shadewatcher: Recommendation-guided cyber threat analysis using
  system audit records.
\newblock In {\em 2022 IEEE Symposium on Security and Privacy (SP)}, pages
  489--506. IEEE, 2022.

\end{thebibliography}

\clearpage

\end{document}